\documentclass[11pt,a4paper]{article}

\usepackage[a4paper, left=1in, right=1in, top=1in, bottom=1in, includehead, includefoot, foot = 0pt, head=0pt]{geometry} 
\usepackage[group-separator={,},group-digits=integer]{siunitx} 
\usepackage{booktabs} 
\usepackage{multirow}
\usepackage{enumitem} 
\usepackage[round,colon,authoryear]{natbib} 

\usepackage{natbib}
\newcommand*{\doi}[1]{\href{http://dx.doi.org/#1}{doi: #1}}
\usepackage[pdftex,colorlinks,citecolor=blue,urlcolor=blue,linkcolor=red]{hyperref} 
\usepackage{graphicx} 
\usepackage{tabularx}
\usepackage{array}
\parskip=\medskipamount 
\usepackage[small,bf,hang]{caption}	
\usepackage{amssymb} 
\usepackage{amsmath} 
\usepackage{bbm} 
\usepackage{amsbsy} 
\usepackage{mathrsfs} 
\usepackage[ruled]{algorithm2e} 
\SetAlCapFnt{\small} 
\SetKw{KwTo}{in}
\usepackage{multirow} 
\usepackage[dvipsnames]{xcolor} 
\usepackage{subcaption} 
\usepackage[hang,flushmargin]{footmisc} 
\usepackage{tikz} 
\usetikzlibrary{arrows,decorations.pathmorphing,decorations.pathreplacing,backgrounds,positioning,fit,petri,patterns}
\usetikzlibrary{matrix, snakes, calc, shapes.misc} 
\tikzstyle{bag} = [align=center]
\usepackage{rotfloat}
\usepackage{titling}
\usepackage{mathtools}
\usepackage[nameinlink]{cleveref}

\usepackage{environ}
\makeatletter
\newsavebox{\measure@tikzpicture}
\NewEnviron{scaletikzpicturetowidth}[1]{%
  \def\tikz@width{#1}%
  \def\tikzscale{1}\begin{lrbox}{\measure@tikzpicture}%
  \BODY
  \end{lrbox}%
  \pgfmathparse{#1/\wd\measure@tikzpicture}%
  \edef\tikzscale{\pgfmathresult}%
  \BODY
}
\makeatother
\usepackage{pgfpages}
\newcommand{\tikzmark}[1]{\tikz[overlay,remember picture] \node (#1) {};}
\newcommand{\DrawBox}[3][]{%
    \tikz[overlay,remember picture]{
    \draw[black,#1]
      ($(#2)+(-0.5em,2.0ex)$) rectangle
      ($(#3)+(0.75em,-0.75ex)$);}
}
\newcommand{\sbt}{\,\begin{picture}(-1,1)(-1,-3)\circle*{3}\end{picture}\ }

\usepackage{soul}
\usepackage{authblk} 

\usepackage{fancyhdr}
\graphicspath{{"../../R code/Images/"}}

\newcommand{\bs}[1]{\boldsymbol{#1}}

\newcommand{\tr}[1]{}

\definecolor{calpolypomonagreen}{rgb}{0.12, 0.3, 0.17}
\usepackage{xr}
\usepackage[normalem]{ulem}

\tikzset{mystripes dist/.initial=1}
\pgfdeclarepatternformonly[/tikz/mystripes dist]{mystripes}
{\pgfpointorigin}{\pgfpoint{1cm}{1cm}}
{\pgfpoint{1cm}{1cm}}
{
  \foreach \x in {0,0.25,0.5,0.75,1}{
    \tikz[overlay]\draw[] (\x, 0) -- ++(1-\x,1-\x);
  }
}

\begin{document}
\title{A hierarchical reserving model for reported non-life insurance claims}

\author[1,3,*]{Jonas Crevecoeur}
\author[1]{Jens Robben}
\author[1,2,3,4]{Katrien Antonio}
\affil[1]{Faculty of Economics and Business, KU Leuven, Belgium.}
\affil[2]{Faculty of Economics and Business, University of Amsterdam, The Netherlands.}
\affil[3]{LRisk, Leuven Research Center on Insurance and Financial Risk Analysis, KU Leuven, Belgium.}
\affil[4]{LStat, Leuven Statistics Research Center, KU Leuven, Belgium.}
\affil[*]{Corresponding author: \href{mailto:jonas.crevecoeur@kuleuven.be}{jonas.crevecoeur@kuleuven.be}, Faculty of Economics and Business, Naamsestraat 69, KU Leuven, Belgium. }
\date{\today}
\maketitle
\thispagestyle{empty}

\begin{abstract}

Traditional non-life reserving models largely neglect the vast amount of information collected over the lifetime of a claim. This information includes covariates describing the policy, claim cause as well as the detailed history collected during a claim's development over time. We present the hierarchical reserving model as a modular framework for integrating a claim's history and claim-specific covariates into the development process. Hierarchical reserving models decompose the joint likelihood of the development process over time. Moreover, they are tailored to the portfolio at hand by adding a layer to the model for each of the events registered during the development of a claim (e.g.~settlement, payment). Layers are modelled with statistical learning (e.g.~generalized linear models) or machine learning methods (e.g.~gradient boosting machines) and use claim-specific covariates. As a result of its flexibility, this framework incorporates many existing reserving models, ranging from aggregate models designed for run-off triangles to individual models using claim-specific covariates. This connection allows us to develop a data-driven strategy for choosing between aggregate and individual reserving; an important decision for reserving practitioners. We illustrate our method with a case study on a real insurance data set and deduce new insights in the covariates driving the development of claims. Moreover, we evaluate the method's performance on a large number of simulated portfolios representing several realistic development scenarios and demonstrate the flexibility and robustness of the hierarchical reserving model.
\end{abstract}



\section{Introduction} \label{section:introduction}

Insurers set aside funds, the so-called reserve, for covering claims from past exposure years. This reserve is often split into a reserve for Incurred, But Not yet Reported (IBNR) claims and a reserve for Reported, But Not yet Settled (RBNS) claims. The specifics of these different reserves should be reflected in the statistical tools used for modelling. Since the claims that compose the IBNR reserve are not yet reported, claim-specific and policy(holder) covariates are unavailable for differentiating the cost per claim. Therefore, IBNR reserving methods mostly put focus on accurately estimating the number of unreported claims, followed by allocating a fixed cost per unreported claim. In RBNS reserving, the insurer is aware of the number of open claims as well as their characteristics and development so-far. This opens the possibility for reserving models that predict the future cost on a per claim basis. This paper puts focus on predicting the RBNS reserve by modelling the individual development of reported claims in the presence of covariates.

Traditionally, the non-life insurance literature has been dominated by analytic models designed for aggregated data, such as the chain ladder method \citep{Mack1993,Mack1999}. These models compress the historical data on the development of claims over time in a two dimensional table, the so-called run-off triangle, by aggregating payments by occurrence and development year. Low data requirements, implementation simplicity and a straightforward interpretation of the predicted reserve justify the popularity of these models. However, by compressing the data valuable insights into the characteristics of individual claims are lost. This makes the reserve less robust against changes in the portfolio composition and extreme one-time events \citep{verdonck2009robustification}. In response to this, individual reserving methods designed for granular data available at the level of individual claims have first been proposed in the nineties. Individual reserving methods remained largely unexplored for about two decades, with revived interest in recent years thanks to an increased focus on the use of big data analytics in insurance.

We identify three streams in the current literature on individual reserving. Following \cite{Norberg1993, Norberg1999}, a first stream analyzes the events registered during a claim's development in continuous time. \cite{lopez2016, lopez2019} adapt regression trees to the right-censoring present in continuous time reserving data. Covariates in these trees capture the heterogeneity in the claim size as well as in the time to settlement of reported claims. In \cite{AntonioPlat2014} hazard rates drive the time to events in the development of claims (e.g.~the time to payment or settlement) and a log-normal regression model is proposed for the payment size. Reserving in continuous time requires a time to event model that allows for multiple types of (recurrent) events. Due to the complexity of the continuous time approach, many individual reserving models are defined in a more convenient discrete time framework, where the events in a claim's lifetime are registered in discrete time periods. Therefore, a second stream of reserving methods models the reserve in discrete time with techniques inspired by insurance pricing. These contributions take advantage of the detailed information on the development of individual claims, as captured by insurance companies. Since these covariates only become available at reporting, such models focus on the reserve for reported, but not settled (RBNS) claims, while using techniques from aggregated reserving to estimate the reserve for unreported claims. \cite{Larsen2007} focuses on Generalized Linear Models (GLMs), \cite{Wuthrich2018} considers regression trees and \cite{Wuthrich2018b} looks at neural networks for reserving. A third stream of papers aggregates the data into multiple run-off triangles. \cite{Martinez2012}, \cite{wahl2019} and \cite{Denuit2016, Denuit2018} consider two, three and four triangles respectively. While the aggregation of the data makes these models easy to implement, covariate information registered for individual claims can not be used.	The recent expansion in (individual) reserving methodology has resulted in a fragmented literature with few comparative studies and no unified approach with proven robustness and general applicability. The lack of a solid modelling framework hinders the implementation of individual reserving in insurance practice. Moreover, only few papers provide data driven guidance on the choice between aggregate and individual reserving, see \citet{arthurmacromicro}, \citet{huang2015stochastic} and \citet{huang2016asymptotic}

We aim to fill this gap in the literature by presenting the hierarchical reserving model as an intuitive framework for RBNS reserving with a focus on applicability in practice. This framework decomposes the joint likelihood of the development of individual claims since reporting in discrete time. Hierarchical reserving models are tailored to the portfolio at hand by adding layers that represent the events (e.g.~settlement, positive or negative payments, changes in the incurred) registered over the lifetime of a claim. This modular approach enables reformulating many existing reserving models as special cases of the hierarchical reserving model, including models based on data aggregated into a run-off triangle. This intuitive model building workflow allows to concentrate on the decisions made during the modelling process, such as model calibration and evaluation. These have received little attention in individual reserving literature up to now with many papers following the model building steps applied in pricing or aggregate reserving. This results in a loss of performance as such methods do not consider the granularity of the data and the presence of censored claim development data.
	
Our hierarchical reserving models are estimated on a European home insurance dataset and their predictive performance is compared with the chain ladder method. In addition, we design a new simulation engine to generate portfolios with the development of individual claims. We evaluate the performance of our proposed reserving framework as well as the performance of the chain ladder method on portfolios simulated along different scenarios (e.g.~presence of an extreme one-time event or faster claim settlement in recent years). Recently, \cite{gabrielli2018individual} and \cite{avanzi2021synthetic} proposed stochastic simulation machines for generating portfolios registering the development of individual claims. \cite{gabrielli2018individual} equip their simulation machine with some policy and claim characteristics, but the portfolios are generated on an annual basis. This makes it more difficult to implement the scenarios that we need for our work using this existing simulation machine. Although the simulation engine of \cite{avanzi2021synthetic} is fully flexible in terms of granularity, it does not generate any claim or policy characteristics other than the occurrence date, reporting date, payment dates and settlement date of the claim. To combine the strengths of both simulation engines, we therefore opt to design our own engine, equipped with some claim and policy characteristics while being in control of the granularity of the data as well as the dependencies among the various model components. 
	
This paper is organized as follows. Section~\ref{section:hrm} introduces the hierarchical reserving model, presents best practices for calibrating this model to insurance data and explains how this model can be used to predict the future RBNS reserve. Section~\ref{section:aggregate} investigates the connection between the hierarchical reserving models proposed at individual claim level and some selected aggregate reserving models from the literature. This results in a data driven strategy for choosing between aggregate and individual reserving. Section~\ref{section:caseStudy} demonstrates this methodology in a case study on a home insurance data set. This is a novel data set, which has not been used before in the literature on reserving. In Section~\ref{section:scenariotesting}, we build two hierarchical reserving models on portfolios simulated along four scenarios generated by our proposed individual claims simulation engine. We compare the predictive performance of the hierarchical reserving model with results obtained with the chain ladder method. Appendix~\ref{appendix:sim.machine} describes the technical details behind the simulation engine. An R package accompanies the paper. This package allows researchers and practitioners to use the simulation engine for the creation of new portfolios from the scenarios discussed in this paper. Next to this, the package makes the implementation of the hierarchical reserving models tailored to their data readily available to researchers and practitioners.

\section{A hierarchical reserving model} \label{section:hrm}
It is common in non-life insurance pricing to decompose the total claim amount on a contract in a frequency and severity contribution \citep{Henckaerts2018}. \cite{Frees2008} extend this idea by proposing a hierarchical insurance pricing model that builds upon three components, namely the frequency, the type and the severity of claims. In this spirit, we propose a hierarchical reserving model, which decomposes the joint likelihood of the claim development process over time and registered events.

\subsection{Model specification} \label{section:model}
We record the development of reported claims in discrete time over a period of $\tau$ years. For each reported claim $k$, $r_k$ denotes the reporting year and the vector $\bs{x}_k$ contains the static claim and policy(holder) information registered at reporting. This paper proposes a so-called hierarchical reserving model with a modular model building approach. Hereto, update vectors $\boldsymbol{U}_k^j$ for $j \geq 1$ describe the claim development information of claim $k$ available at the end of development year $j$ since reporting, i.e.~at the end of calendar year $r_k + j - 1$. The update vector $\boldsymbol{U}_k^1$ thus denotes the available claim information at the end of the year in which the claim was reported. The length and components of $\boldsymbol{U}_k^{j}$ are tailored to the events registered in the portfolio at hand (e.g. claim settlement, change in the incurred, involvement of a lawyer). The resulting hierarchical reserving model is built on two fundamental assumptions.

\paragraph{Model assumptions}
\begin{itemize}
	\item[(A1)] All claims settle within $d$ years after reporting.
	\item[(A2)] The development of a claim is independent of the development of the other claims in the portfolio.
\end{itemize}

Although upper limit $d$ on the settlement delay is not necessarily limited to the length of the observation window, we implicitly assume $d = \tau$ for notational convenience. Given these assumptions, the set of observed claim updates after reporting for a portfolio of $n$ claims is
\begin{equation*}
	\mathcal{R}^{\texttt{Obs}} = \{\bs{U}_{k}^{j} \mid k = 1 ,\ldots,n, j = 1, \ldots, \tau_k \},
\end{equation*} 
with $\tau_k = \min(d, \tau - r_k + 1)$ the number of observed development years for claim $k$. The associated likelihood is
\begin{equation*}
	\mathcal{L}\left(\mathcal{R}^{\texttt{Obs}} \mid \boldsymbol{X} \right) = \prod_{k = 1}^n f\left(\bs{U}^{1}_k, \ldots, \bs{U}^{\tau_k}_k \mid \bs{x}_k\right),
\end{equation*}
where $\boldsymbol{X}$ is the observed matrix of row vectors $\boldsymbol{x_k}$ and assumption (A2) is used to write the likelihood as a product of claim-specific likelihood contributions. Inspired by \cite{Frees2008}, we introduce a hierarchical structure in this likelihood by applying the law of conditional probability twice. First, we include the temporal dimension and split the likelihood by development year in chronological order
\begin{align*}
	\mathcal{L}\left( \mathcal{R}^{\texttt{Obs}} \mid \boldsymbol{X} \right) &= \prod_{k = 1}^n \prod_{j=1}^{\tau_k} f\left(\bs{U}_k^{j} \mid \bs{U}_k^{1}, \ldots, \bs{U}_k^{j-1}, \bs{x}_k\right).								
\end{align*}
By conditioning on past events, we acknowledge that the future development of a claim depends on its development in previous years. Second, we split the likelihood by iterating over the elements of the vector $\bs{U}_k^j$:
\begin{align}
\mathcal{L}\left( \mathcal{R}^{\texttt{Obs}} \mid \boldsymbol{X} \right) &=  \prod_{k = 1}^n \prod_{j=1}^{\tau_k} \prod_{l=1}^{s} f\left(U_{k,l}^{j} \mid \bs{U}_k^{1}, \ldots, \bs{U}_k^{j-1}, U_{k,1}^{j}, \ldots, U_{k,l-1}^{j}, \bs{x}_k\right), \label{eq:likelihood_general}
\end{align}
where $s$ is the length of the update vector $\bs{U}_k^{j}$. The elements $U_{k,l}^{j}$ for $l = 1,\: ..., \: s$ of the vector $\bs{U}_k^j$ correspond to the different events registered for claim $k$ in development year $j$ since reporting. In the remainder of this paper, we refer to these $s$ events as the layers of the hierarchical model. The order of the layers is an important model choice, since the outcome of a layer becomes a covariate when modelling higher indexed layers. Because assumptions (A1-A2) are common in reserving literature, many discrete time reserving models can be seen as a special case of our hierarchical reserving framework. Notice that in contrast to the chain ladder method, the hierarchical framework includes the entire history of the claim and thus allows for non-Markovian models.

When we apply the hierarchical claim development model to a specific portfolio, we extend the assumptions (A1-A2) with an additional assumption that tailors the structure of the update vector $\bs{U}_k^{j}$ to the portfolio at hand. For example, in the case study in Section~\ref{section:caseStudy} and the simulation study performed in Section~\ref{section:scenariotesting}, we specify a three-layer hierarchical model for $\bs{U}_k^{j}$ with the distributional assumptions listed below in (A3).

\paragraph{Example of layers in the hierarchical reserving model and their distributional assumptions}
\begin{itemize}
	\item[(A3)] The update vector $\bs{U}_k^{j}$ for claim $k$ in development year $j$ since reporting has three layers \\ $\bs{U}_k^{j} = (U_{k, 1}^{j}, U_{k, 2}^{j}, U_{k, 3}^{j}) = (C_k^{j}, P_k^{j}, Y_k^{j})$, where
	\begin{itemize}
		\item[$\sbt$] $C_k^{j}$ is the settlement indicator which is one when claim $k$ settles in development year $j$ and zero otherwise. Conditional on past events, the settlement indicator follows a Bernoulli distribution with
		$$C_k^{j} \mid_{\bs{U}_k^{1}, \ldots, \bs{U}_k^{j-1}, \bs{x}_k}  \sim \texttt{Bernoulli}\left( p\left( \bs{U}_k^{1}, \ldots, \bs{U}_k^{j-1}, \bs{x}_k\right) \right),$$
		where $p(\cdot)$ denotes the unknown settlement probability function.
		\item[$\sbt$] $P_k^{j}$ is the payment indicator which is one when there is a payment for claim $k$ in development year $j$ and zero otherwise. Conditional on past events, the payment indicator follows a Bernoulli distribution with
		$$P_k^{j} \mid_{\bs{U}_k^{1}, \ldots, \bs{U}_k^{j-1}, C_k^{j}, \bs{x}_k}  \sim \texttt{Bernoulli}\left( q\left( \bs{U}_k^{1}, \ldots, \bs{U}_k^{j-1}, C_k^{j}, \bs{x}_k \right) \right),$$
		where $q(\cdot)$ denotes the unknown payment probability function.
		\item[$\sbt$] $Y_k^{j}$ is the payment size, given that there was a payment in development year $j$. Conditional on past events, the payment size is gamma distributed with mean
		$$E\left(Y_k^{j} \mid \bs{U}_k^{1}, \ldots, \bs{U}_k^{j-1}, C_k^{j}, P_k^{j}, \bs{x}_k\right) = \mu\left(\bs{U}_k^{1}, \ldots, \bs{U}_k^{j-1}, C_k^{j}, P_k^{j}, \bs{x}_k\right) $$
		and variance
		$$ \sigma^2\left(Y_k^{j} \mid \bs{U}_k^{1}, \ldots, \bs{U}_k^{j-1}, C_k^{j}, P_k^{j}, \bs{x}_k\right) = \theta \cdot \mu\left(\bs{U}_k^{1}, \ldots, \bs{U}_k^{j-1}, C_k^{j}, P_k^{j}, \bs{x}_k\right).$$
	\end{itemize} 
\end{itemize}

In the above example, we structure the development of claims with a simple three-layer hierarchical model. Conditioning on the settlement status in past years allows us to train the model on the development of open claims only, whereas without settlement indicator the model would potentially predict new payments for already settled claims. Moreover, by choosing settlement as the first layer in the hierarchical model, settlement becomes a covariate when modelling higher indexed layers. The gamma distribution for the sizes is frequently used in the insurance pricing literature to model attritional losses \citep{Henckaerts2021}. Choosing a strictly positive distribution assumes that there are no recoveries in the portfolio. In portfolios in which recoveries are common, additional layers should be added to the hierarchical model to allow for negative payments.  

\subsection{Calibration} \label{section:covariateselection}

The individual layers in the hierarchical reserving model can be modelled with any predictive modelling technique. The case study in Section~\ref{section:caseStudy} and the simulation study in Section~\ref{section:scenariotesting} calibrate both a Generalized Linear Model (GLM) as well as a Gradient Boosting Model (GBM) to the layers outlined in (A3). Although standard procedures are available to calibrate these models, special attention is required for the variable selection process or (hyper) parameter tuning steps. In reserving, the historical, observed data mainly contain records from early development years, whereas the future predictions are more oriented towards later development years. This imbalance between the training and prediction data set\footnote{The prediction data set consists of all update vectors $\bs{U}_k^j$ that are not yet observed at the evaluation date, i.e.~the set $\mathcal{R}^{\texttt{Pred}} = \{\bs{U}_{k}^{j} \mid k = 1 ,\ldots,n, j = \min(\tau_k + 1,d), \ldots, d \}$.} poses a model risk when covariates exhibit a different effect on the first versus the later development years. In machine learning literature this phenomenon is known as a covariate shift \citep{Sugiyama2007}. Following \cite{Sugiyama2007b}, we correct for a potential covariate shift by maximizing a weighted likelihood, i.e.
\begin{align}
\mathcal{L}^{\texttt{weighted} }\left( \mathcal{R}^{\texttt{Obs}} \mid \boldsymbol{X} \right) &=  \prod_{k = 1}^n \prod_{j=1}^{\tau_k} w_{j} \prod_{l=1}^{s} f\left(U_{k,l}^{j} \mid \bs{U}_k^{1}, \ldots, \bs{U}_k^{j-1}, U_{k,1}^{j}, \ldots, U_{k,l-1}^{j}, \bs{x}_k\right), \label{eq:likelihood_general_weighted}
\end{align}
where $w_j$ is the weight assigned to an observation from development year $j$ since reporting. Following \cite{Sugiyama2007b}, we define these weights as the ratio of the number of records from development year $j$ in the prediction data set to the number of records from development year $j$ in the training data set. For typical reserving data sets this ratio is observed and can be computed as
\begin{align*}
w_1 = 0, \hspace{0.5cm} w_j = \frac{\sum_{i=d-j+2}^{d} n_i}{\sum_{i=1}^{d-j+1} n_i}, \hspace{0.5cm}  &j = 2, 3, \ldots, d,
\end{align*}
where $n_i$ is the number of reported claims in reporting year $i$. These weights increase in $j$ and assign more weight to observations from later development years. A weight of zero is assigned to observations from the first development year since reporting, as none of these observations are included in the prediction data set. As a consequence, reserving models only focus on later development years ($j \geq 2$) during the calibration process. This is advantageous since the claim development process in the first development year often deviates significantly from later development years. A disadvantage is that the zero weight prevents the estimation of covariate effects that are only observed in the most recent reporting year. A relevant example of this limitation is the inclusion of \texttt{rep.year} as a categorical covariate in the reserving model. When $w_1 = 0$ a separate parameter can not be calibrated for the level corresponding to the most recent reporting year. In the case study and simulation study we overcome this disadvantage by assigning a small positive weight to observations from development year one, i.e. $w_1 = 10^{-6}$. Alternatively, the levels corresponding to the two most recent reporting years in the categorical covariate \texttt{rep.year} can be merged.

When selecting covariates or tuning (hyper) parameters, we maximize \eqref{eq:likelihood_general_weighted} in a $5$-fold cross-validation scheme. Hereto we calibrate predictive models per layer $l$ and allocate observations at the level of a claim $k$ and a development year $j$ (see \eqref{eq:likelihood_general_weighted}) to different folds. We then re-estimate the model $f(\cdot)$ using the optimal set of covariates or tuning parameters by maximizing the weighted likelihood in Equation~\eqref{eq:likelihood_general_weighted} on the entire training data set. \cite{Sugiyama2007b} point out that consistent parameter estimators result and that the explained weighted cross-validation is almost unbiased, even under a covariate shift.

\subsection{Restoring the balance property}\label{subsec:balanceprop}
After the calibration of the hierarchical reserving model, we eliminate the portfolio bias in the layer-specific parameter estimates, before we proceed to the actual prediction of the RBNS reserves. This so-called bias correction step is inspired by \cite{wuthrich2020bias} and \cite{gabrielli2021individual}. GLMs with canonical link lead to unbiased estimates at portfolio level (see formula (10) in \citet{nelder1972generalized}), also known as the balance property. As such, they do not require a bias correction step.

\cite{wuthrich2020bias} proposes to restore the balance property in neural network models subject to early stopping rules by introducing an additional GLM step. \cite{gabrielli2021individual} estimates development year specific variance parameters in the assumed log-normal distribution for the payment sizes in such a way that the balance property holds. In this paper, we define development year specific bias correction factors $b_l^j$ for each layer $l$ (with $l = 1,\: \ldots,\: s$):
\begin{align*}
b_l^j = \dfrac{\displaystyle \sum_{k = 1}^n U_{k,l}^j}{\displaystyle \sum_{k = 1}^n \hat{U}_{k,l}^j}, \hspace{0.5cm}  &j = 1,2, \ldots, \tau,
\end{align*}
where the predictions $\hat{U}_{k,l}^j$ are defined as
\begin{align*}
\hat{U}_{k,l}^j := \hat{\text{E}}\left(U_{k,l}^{j} \mid \bs{U}_k^{1}, \ldots, \bs{U}_k^{j-1}, U_{k,1}^{j}, \ldots, U_{k,l-1}^{j}, \bs{x}_k\right), 
\end{align*}
and where $\hat{\text{E}}(\cdot)$ denotes the predicted mean. To restore the balance property, we multiply the in-sample predictions $\hat{U}_{k,l}^j$ with these development year specific bias correction factors $b_{l}^{j}$. In rare occasions where such a correction would result in a response outside the valid domain (e.g.~a probability exceeding one), we replace the outcome by the corresponding boundary value. 

\subsection{Predicting the future development of claims}  \label{section:grm_prediction}
Algorithm~\ref{algorithm:simulation} simulates the development of a reported claim $k$ beyond its observed development window $\tau_k$. In line with the hierarchical structure of the model, development years since reporting are simulated in chronological order and within a development year, the simulation algorithm respects the order of the layers. This order is important, since simulated values from previous development years and lower indexed layers become inputs for later development years and higher indexed layers. 

\begin{algorithm}[H]
	
    \SetAlgoLined
    \KwIn{the observed development of reported claims}
    \KwOut{simulation of the future development of reported claims}
        \ForEach{\text{claim} $k$}{
        	\For{\text{development year} $j$ \KwTo{} $\tau_k+1, \ldots, d$ }{
            	\For{\text{hierarchical layer} $l$ \KwTo{} $1, \ldots, s$ }{
            	\begin{footnotesize}
            	Compute the out-of-sample prediction: $\hat{U}_{k,l}^j := \hat{\text{E}}\left(U_{k,l}^{j} \mid \bs{U}_k^{1}, \ldots, \bs{U}_k^{j-1}, U_{k,1}^{j}, \ldots, U_{k,l-1}^{j}, \bs{x}_k\right)$ \\ 
            	Apply bias correction step: redefine  $\hat{U}_{k,l}^j := \hat{U}_{k,l}^j \cdot b_l^j$ \\
            	Simulate $U_{k,l}^{j}$ from the estimated distribution with mean $\hat{U}_{k,l}^j$ for the layer $l$ outcome variable\footnote{Other parameters in the imposed distribution for the layer $l$ outcome variable are estimated using maximum likelihood.}
            	\end{footnotesize}
        		}
        	}    
        }
    \caption{Simulating the future development of reported claims}
    \label{algorithm:simulation}
\end{algorithm}

Following this algorithm, the simulated data set has the same hierarchical layered structure as the input data set, which enables the calculation of aggregated quantities for the events registered in the update vector $\boldsymbol{U}_k^{j}$. For example, given the specific hierarchical structure in assumption (A3), we obtain estimates for the number of open claims, the number of payments and the total payment size. Prediction intervals for these reserving quantities are obtained by generating multiple paths with Algorithm~\ref{algorithm:simulation}. 

\subsection{Implementation in R} A dedicated \texttt{R} package called \texttt{hirem} \citep{CrevecoeurRobben2021} is available to calibrate the hierarchical reserving models and to simulate the future development of claims. In this package, layer-specific outcome variables can be modelled with generalized linear models (GLMs) or gradient boosting models (GBMs). We hereby rely on the implementation from \cite{Southworth2015} of the \texttt{gbm} package, which adds the gamma loss function to the original package developed by \cite{Greenwell2018}.

\section{Bridging aggregate and individual reserving methods} \label{section:aggregate}
Most claim reserving models proposed in actuarial literature are formulated for data aggregated into run-off triangles. In this section we start from the data registered at the level of individual claims and illustrate how aggregate reserving models can be retrieved as special cases of the hierarchical reserving model. Section~\ref{section:aggregate_1} investigates the simplified case of a hierarchical model with independent layers. Section~\ref{section:aggregate_2} then enables a simple, but widely used dependency structure between the layers. The results of these sections offer valuable insights and statistical tools for choosing between aggregate and individual reserving. Section~\ref{section:aggregate_3} constructs some hierarchical reserving models that are directly inspired by recent literature contributions on aggregate reserving with multiple run-off triangles. The hierarchical reserving model offers new insights on possible extensions of these aggregate models to the level of individual claims. Moreover, as a unified framework for RBNS reserving this facilitates model comparison, calibration and the generation of future paths of a claim's development.

\subsection{From individual hierarchical reserving models with independent layers to aggregate reserving models} \label{section:aggregate_1}
We start from individual claims data structured in update vectors $\boldsymbol{U}^j_k$, as described in Section~\ref{section:model}. For each layer $l$ in the update vector, we construct a run-off triangle $(X_{l}^{ij})_{1\leq i, j\leq d}$ with cells
$$X_{l}^{ij} = \sum_{k: r_k = i} U_{k, l}^{j}.$$

In line with our focus on modelling RBNS claims, we aggregate by reporting year (denoted with $i$) and development year since reporting (denoted with $j$) instead of the traditional set-up where aggregation goes per occurrence year and development year since occurrence. As such, we model the development of claims since reporting. Although we denote the reporting year of claim $k$ by $r_k$, we keep the traditional index $i$ for the rows, i.e.~the reporting years, in the constructed run-off triangles.

Let us now assume that the individual updates $U_{k,l}^j$ in layer $l$ depend multiplicatively on the reporting year and the development year since reporting, i.e.
\begin{equation}
\text{E}\left(U_{k, l}^{j} \mid \boldsymbol{U}_k^{1}, \ldots,\boldsymbol{U}_k^{j-1} , U^j_{k,1}, \ldots, U^j_{k,l-1}, \boldsymbol{x}_k \right) = \alpha_{r_k, l}  \cdot \beta_{j, l} \quad \text{with}\quad \sum_{j=1}^{d} \beta_{j, l} = 1, \label{eq:aggregate_1}
\end{equation}
where $\alpha_{r_k, l}$ is the layer $l$ specific effect of reporting year $r_k$ and $\beta_{j, l}$ is the effect of development year $j$. When we aggregate these individual updates into a run-off triangle, the resulting cell values follow a similar multiplicative structure,\footnote{For notational purposes, we omit the dependence on the individual claim updates $\boldsymbol{U}_k^{1}, \ldots,\boldsymbol{U}_k^{j-1} , U^j_{k,1}, \ldots, U^j_{k,l-1}$ and on the claim information $\boldsymbol{x}_k$ in the remaining of this section as well as in Sections~\ref{section:aggregate_2} and \ref{section:aggregate_3}.} i.e.
\begin{equation}
\begin{aligned}
\text{E}(X_{l}^{ij}) = \text{E}\left(\sum_{k: r_k = i} U_{k, l}^{j}  \right) = n_i \cdot \alpha_{i, l} \cdot \beta_{j, l} \coloneqq  \tilde{\alpha}_{i, l} \cdot \beta_{j, l}, \label{eq:aggregate_0}
\end{aligned}
\end{equation}
where $n_i$ is the observed number of reported claims in reporting year $i$. As a result, we can calibrate individual hierarchical reserving models that only depend multiplicatively on the reporting and development year using data aggregated into run-off triangles.

Matching \eqref{eq:aggregate_1} with the original hierarchical reserving model specification in \eqref{eq:likelihood_general}, we rephrase the expected value of the update vector $\boldsymbol{U}_{k}^{j}$ at the individual level in full generality as
\begin{equation*}
\begin{aligned}
\text{E}\Big(U_{k, l}^{j} \mid \boldsymbol{U}_k^{1}, \ldots, &\boldsymbol{U}_k^{j-1} , U^j_{k,1}, \ldots, U^j_{k,l-1}, \boldsymbol{x}_k \Big) \\
&=\alpha_{r_{k}, l} \cdot \beta_{j, l}\cdot \phi_l \Big(\boldsymbol{U}_k^{1}, \ldots, \boldsymbol{U}_k^{j-1} , U^j_{k,1}, \ldots, U^j_{k,l-1}, \boldsymbol{x}_k \Big),
\end{aligned}
\end{equation*}
where $\phi_l(\cdot)$ represents the effect of all other covariates on the outcome variable of layer $l$. When we add a distributional assumption for $\bs{U}_{k}^{j}$, testing for $\phi_l(\cdot) = 1$ (for each layer $l$) will provide guidance in the choice between an aggregate and individual reserving model formulation. Since the models with and without $\phi_l(\cdot)$ are nested, a likelihood ratio test can be used. 

\subsection{From individual hierarchical reserving models with dependent layers to aggregate reserving models} \label{section:aggregate_2}
The reserving models constructed in Section~\ref{section:aggregate_1} treat each layer independently of the others. This results in simple aggregated models, where each layer is estimated from a single run-off triangle, independent from the other layers. However, in most multi-layer hierarchical structures, some dependence between the layers is inevitable and offering a simple framework to include these dependencies is one of the main motivations for the hierarchical reserving model. By means of example, this section investigates the special, but common setting of a two-layer hierarchical model in which layer one is a binary random variable and layer two is zero whenever layer one equals zero. As an example, $U_{k,1}^j$ indicates whether a payment is made in development year $j$ since reporting for claim $k$. The corresponding payment size, in case of a payment, is then denoted by $U_{k,2}^j$. When there is no payment, the payment size is zero.

Again focusing on a multiplicative structure in reporting and development year, we specify an example of such a two-layer hierarchical model:
\begin{align*}
	\text{E}(U_{k, 1}^{j}) &=  \alpha_{r_{k}, 1} \cdot \beta_{j, 1} \cdot \phi_1 \left(\boldsymbol{U}_k^{1}, \ldots, \boldsymbol{U}_k^{j-1}, \boldsymbol{x}_k \right) \\
	\text{E}(U_{k, 2}^{j}) &= \begin{cases}
		\alpha_{r_{k}, 2} \cdot \beta_{j, 2} \cdot \phi_2\left(\boldsymbol{U}_k^{1}, \ldots, \boldsymbol{U}_k^{j-1}, \boldsymbol{x}_k \right) & \text{ if } \: U_{k, 1}^{j} = 1 \\
		0 & \text{ if } \: U_{k, 1}^{j} = 0
	\end{cases}, 
\end{align*}
where $\sum_{j=1}^{d} \beta_{j, l} = 1$ for $l \in \{1, 2\}$. We again omit the dependence on the individual claim updates and on the static claim information vector within the expectation for notational convenience. When $\phi_1(\cdot)$ and $\phi_2(\cdot)$ are both equal to one, the claim development pattern depends only on the reporting year and the development year since reporting in a multiplicative way. At aggregate level we then retrieve
\begin{align}
	\text{E}(X_{1}^{ij}) = \text{E}\left(\sum_{k: r_k = i} U_{k, 1}^{j}\right) &= n_i \cdot \alpha_{i, 1} \cdot \beta_{j, 1} \coloneqq \tilde{\alpha}_{i, 1} \cdot \beta_{j, 1} \notag\\
	\text{E}(X_{2}^{ij} \mid X_{1}^{ij}) = \text{E}\left(\sum_{{k: r_k = i}} U_{k, 2}^{j}\right) &= X_1^{ij} \cdot \alpha_{i, 2} \cdot \beta_{j, 2}. \label{eq:aggregated_model2}
\end{align}
Here, $X_1^{ij}$ captures the number of payments made in development year $j$ since reporting for those claims in the portfolio reported in year $i$. The total of the corresponding payment sizes is denoted by $X_2^{ij}$. When we calibrate the model for the second layer, the cells in the observed upper triangle corresponding to the first layer act as an exposure term. When estimating the future reserve, this exposure term, $X_1^{ij}$, should be estimated using the model proposed for the first layer. In the example, the number of payments then becomes the exposure for the total payment size. Similar to Section~\ref{section:aggregate_1}, statistical tests for $ \phi_1(\cdot) = \phi_2(\cdot) = 1$ offer data driven tools for choosing between an individual and an aggregate reserving strategy.

\subsection{Hierarchical reserving models inspired by aggregate reserving models proposed for multiple run-off triangles} \label{section:aggregate_3}
As an illustration of the generality of our framework, we construct hierarchical reserving models inspired by recent contributions on aggregate reserving models with data structured in multiple triangles. We discuss two examples of such models, namely the double chain ladder \citep{Martinez2012} and the collective reserving model \citep{wahl2019}. As motivated in Section~\ref{section:introduction}, we limit our analysis to the RBNS part of these aggregate models.

\paragraph{Double chain ladder} The double chain ladder (DCL) \citep{Martinez2012} extends the chain ladder method to two run-off triangles and constructs separate estimates for the IBNR and RBNS reserve. Since we only consider the development of claims after reporting, we focus on the triangle of claim payment sizes and construct a one-layer hierarchical model. DCL structures the expected payment size for a claim $k$ in development year $j$ since reporting, denoted $U^{j}_{k, 1}$, as
$$ \text{E}(U_{k,1}^{j}) = \tilde{\pi}_j \cdot \tilde{\mu}_j \cdot \gamma_{i_k},$$
where $i_k$ denotes the occurrence year of the claim and $\tilde{\pi}_j$ and $\tilde{\mu}_j$ refer to the payment probability and the average payment size in development year $j$ since reporting, respectively. The coefficient $\gamma_{i_k}$ adjusts the size of the payments from occurrence year $i_k$ for inflation. Letting inflation depend on the occurrence year is natural in DCL, which aggregates run-off triangles by occurrence year and development year since occurrence. Since run-off triangles based on the hierarchical reserving model aggregate by reporting year and development year since reporting, it is in our framework more natural to model inflation per reporting year. Therefore we change the occurrence year effect $\gamma_{i_k}$ by a reporting year effect $\gamma_{r_k}$. The expected payment size for a claim $k$ in development year $j$ since reporting then becomes 
$$ \text{E}(U_{k,1}^{j}) = \tilde{\pi}_j \cdot \tilde{\mu}_j \cdot \gamma_{r_k}. $$
By aggregating these expected payment sizes by reporting year $i$ and development year $j$ since reporting, we retrieve
$$ \text{E}(X_{1}^{ij}) = \text{E}\left( \sum_{k: r_k = i} U_{k,1}^{j} \right) = n_i \cdot \tilde{\pi}_j \cdot \tilde{\mu}_j \cdot \gamma_{i}.$$
This is the same model as \eqref{eq:aggregate_1}, when we rewrite $\alpha_i = n_i \cdot \gamma_{i}$ and $\beta_j = \tilde{\pi}_j \cdot \tilde{\mu}_j$.

\paragraph{The collective reserving model} Extending the earlier work of \cite{Verrall2010} and \cite{Martinez2012}, the collective reserving model \citep{wahl2019} structures the claim development after reporting in two layers. These layers represent the number of payments and the size per payment. Inspired by \cite{wahl2019}'s aggregate model, we specify a hierarchical reserving model that structures the individual updates as
\begin{align*}
	U_{k, 1}^{j} &\sim \texttt{Poisson}(\lambda_j) \tag{\text{number of payments}}, \\
	\text{E}(U_{k, 2}^{j} \mid U_{k, 1}^{j}) &= \mu(i_k, r_k, j) \cdot U_{k, 1}^{j}, \tag{\text{payment size}}
\end{align*}
Here, a claim can have multiple payments in the same year $j$ since reporting. The number of payments that are made for claim $k$ in development year $j$ since reporting, denoted by $U_{k,1}^j$, is modelled with a Poisson distribution. The average size of a payment is $\mu(i_k, r_k, j)$, which depends on the claim's occurrence year, reporting year and development year since reporting. When $\mu(i_k, r_k, j) = \alpha_{r_k} \cdot \beta_j$, we aggregate the model by reporting year $i$ and development year $j$ since reporting to
\begin{align*}
	\text{E}(X^{ij}_1) &= n_i \cdot \lambda_j, \\
	\text{E}(X^{ij}_2 \mid X_1^{ij}) &= X^{ij}_1 \cdot \alpha_i \cdot \beta_j,
\end{align*}
where $n_i$ denotes the number of claims reported in reporting year $i$. This representation is almost identical to \eqref{eq:aggregated_model2}, with the distinction that the estimated effect of reporting year $i$ for the number of payments is now replaced by the observed claim count $n_i$.

\section{Case study: European home insurance portfolio} \label{section:caseStudy}
This case study models the RBNS reserve for a European home insurance portfolio. This insurance product reimburses damages to the insured property and its contents resulting from a wide range of causes including fire damage, water damage and theft. For reasons of confidentiality we can not disclose the size of the portfolio and the associated reserve. Therefore, we express the performance of the investigated reserving methods via a percentage error measure, comparing the actual and the predicted reserve in some out-of-time evaluations.

\subsection{Data characteristics}
\begin{figure}[ht!]
\centering
\includegraphics[width = 0.9\textwidth]{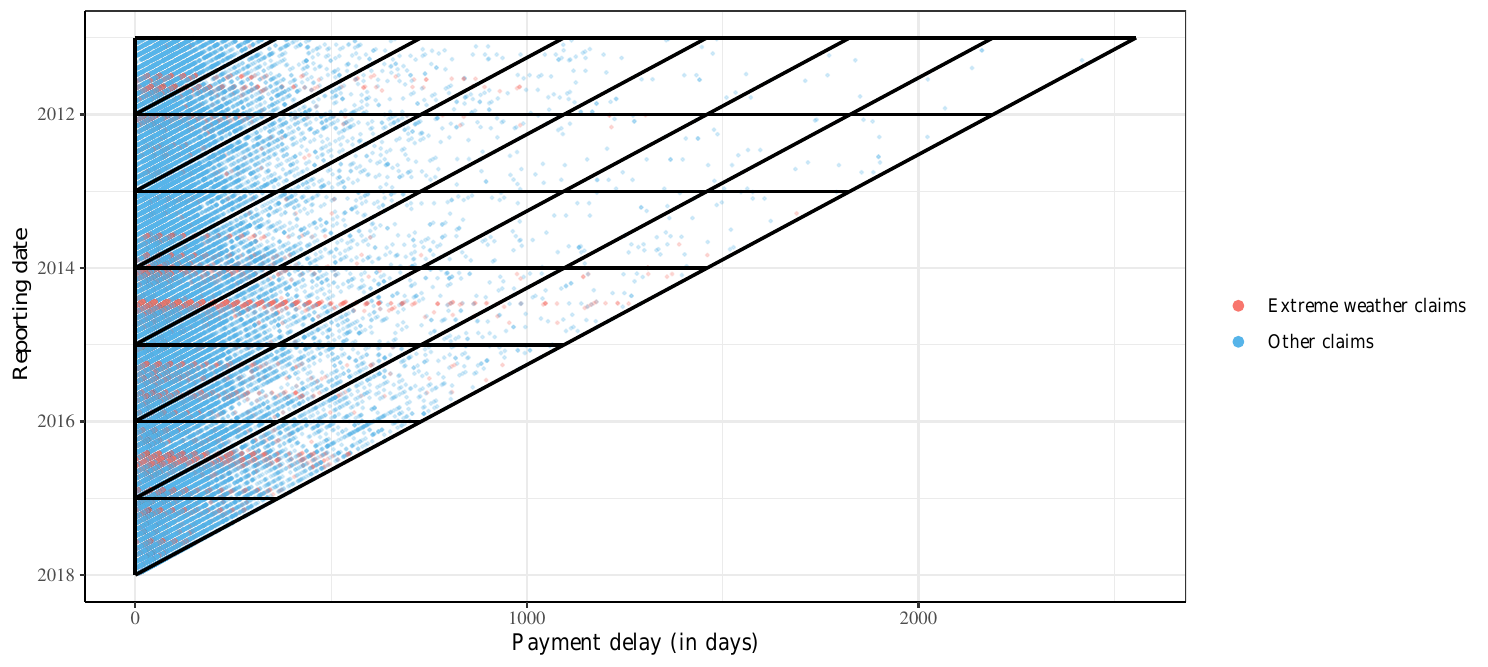}
\caption{Payments structured by reporting date and payment delay in days. Every dot represents a single payment and one claim can have multiple payments. A grid indicates how individual payments would be aggregated when constructing a yearly run-off triangle. Claims resulting from extreme weather (e.g.~a storm) are coloured in red.}
\label{figure:individualTriangle}
\end{figure}

We observe the development of individual claims over a seven year period from January 2011 until December 2017. Figure~\ref{figure:individualTriangle} structures individual payments by the reporting date of the claim (vertical axis) and the number of days elapsed since reporting (horizontal axis). Every dot represents a single payment and one claim can have multiple payments. A triangular structure appears, since the claim development after December, 2017 is censored. Home insurance is a short tailed business line, with many payments in the first years after reporting. The black grid in Figure~\ref{figure:individualTriangle} visualizes the aggregation of individual payments when constructing a yearly run-off triangle. As shown in this triangle, extreme weather events cause sudden spikes in the number of reported claims. This has a large impact on the stability of the run-off triangle in traditional, aggregate reserving. Therefore, insurers most often reserve these claims separately based on expert opinion. In this paper we analyze the robustness of various hierarchical reserving methods by predicting the future reserve with and without extreme weather events. Table~\ref{table:covariates} provides a detailed description of the available covariates. We group these covariates into four categories. Policy covariates identify the policy or policyholder entering the claim. These covariates are available when pricing the contract. Claim covariates describe the static characteristics of the claim. These covariates become available in the reporting year of the claim. Development covariates describe the yearly evolution of the claim. The layer outcome variables and covariates define the layers of the hierarchical reserving model.

\begin{table}[ht!]
\centering
\begin{tabular}{l l}
\toprule
\multicolumn{2}{l}{{\bf Policy covariates}} \\
\texttt{valuables} & Objects were declared with a value exceeding the standard cover: yes or no \\
\texttt{age.insured} & The age of the policyholder \\
\texttt{profession} & Profession of the policyholder, 16 categories \\
\texttt{sex} & Gender of the policyholder: male or female \\
\texttt{construction.year} & The year in which the building was constructed \\
\texttt{property.value} & The value of the property in Euro \\
\multicolumn{2}{l}{{\bf Claim covariates}} \\
\texttt{occ.date} & Date on which the accident occurred \\
\texttt{rep.date} & Date on which the claim was reported to the insurer \\
\texttt{rep.delay} & Delay in days between the occurrence and reporting of the claim \\
\texttt{rep.year} & Calendar year in which the claim was reported \\
\texttt{rep.month} & Calendar month in which the claim was reported (Jan - Dec)  \\
\texttt{coverage} & The main coverage applicable to the claim: theft, building or contents \\
\texttt{catnat} & The cause of the claim, grouped in 12 categories \\
\texttt{extreme.weather} & Claim is the result of extreme weather (e.g.~storm): yes or no \\
\texttt{initial.reserve} & Expert estimate of the initial reserve at the end of the reporting year \\
\multicolumn{2}{l}{{\bf Development covariates}} \\ 
\texttt{dev.year} & The number of years elapsed since the reporting of the claim \\
\texttt{calendar.year} & Number of years elapsed between the start of the portfolio and \texttt{dev.year} \\
\multicolumn{2}{l}{{\bf Layer outcome variables}} \\ 
\texttt{settlement} & The claim settles in the current development year: yes or no \\
\texttt{payment} & A payment occurs in the current development year: yes or no \\
\texttt{size} & Total amount paid in the current development year \\
\bottomrule
\end{tabular}
\caption{List of covariates available in the home insurance data set.  A level NA (not available) identifies the records with no registered value for a covariate.}
\label{table:covariates}
\end{table}

\begin{figure}[ht!]
\centering
\includegraphics[width = 0.8\textwidth]{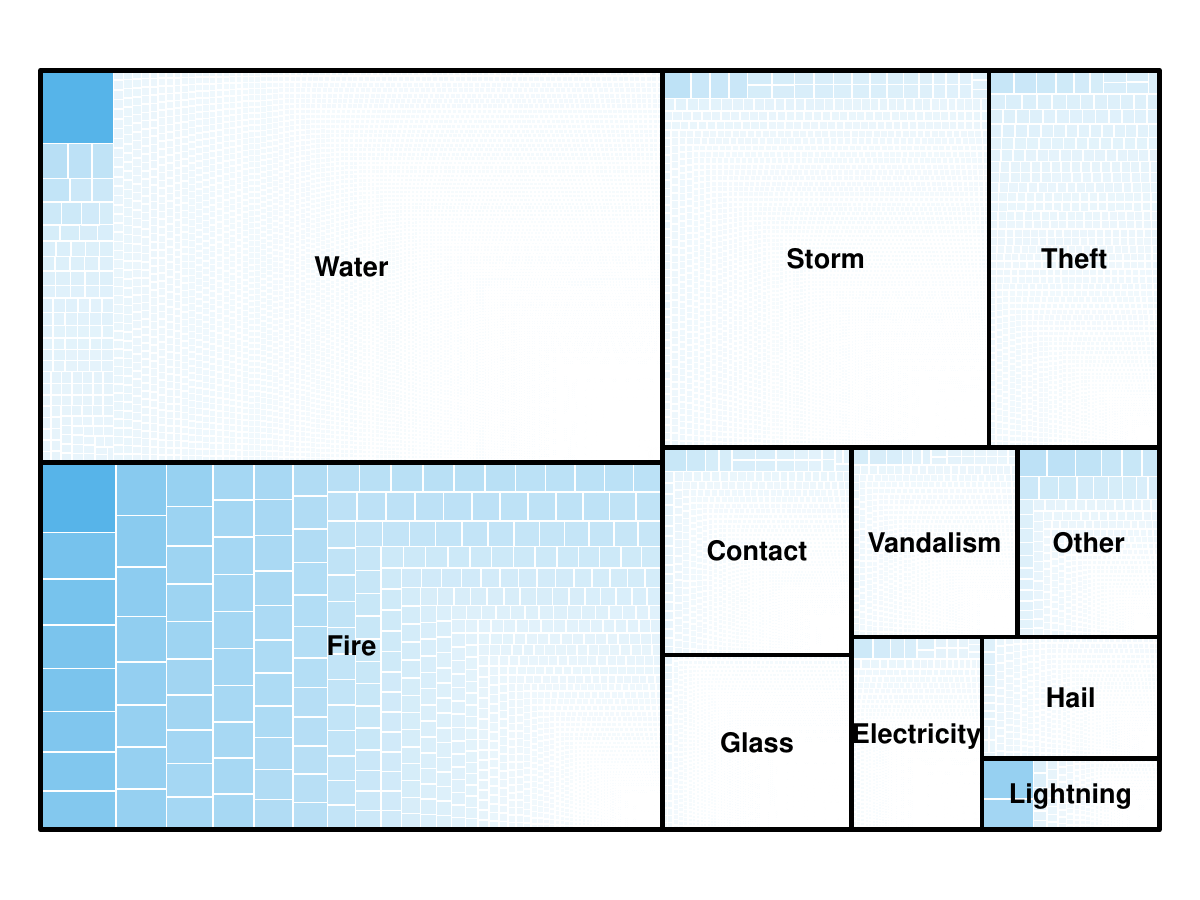}
\caption{Treemap of individual claims observed on 31 December, 2017 grouped into the 12 risk categories present in the portfolio as coded in the covariate $\texttt{catnat}$. Each claim is represented by a rectangle, where the size of this rectangle visualizes the amount paid for that claim by the end of December, 2017.}
\label{figure:treemap}
\end{figure}

In Figure~\ref{figure:treemap} a treemap visualizes the available claims grouped into the 12 risk categories as coded in the covariate \texttt{catnat}. Each claim is represented by a rectangle, where the size of this rectangle visualizes the amount paid for that claim by the end of December, 2017. Water and fire damage are the most important insurance covers in this portfolio. Together these risks generate more than half of the total claim cost. Fire claims are typically larger than non-fire claims. Although less than $5\%$ of all claims are related to fire, these claims represent more than $25\%$ of the total cost. The large difference between the average size of fire claims versus the average size of non-fire claims, motivates us to build separate reserving models for fire claims on the one hand and non-fire claims on the other hand. Estimating separate reserves for risks with a different development pattern is a common approach in traditional reserving. Alternatively, we can distinguish fire and non-fire claims by including a covariate in the hierarchical reserving model. However, the latter approach would result in an unfair comparison between individual models, which can use this covariate, and traditional reserving methods for aggregate data, which can not use this covariate.

\subsection{Hierarchical reserving models for fire and non-fire claims}

We analyse the performance of hierarchical reserving models based on GLMs, GBMs and the chain ladder method on 365 evaluation dates between January 1, 2015 and December 31, 2015. Instead of a single out-of-time evaluation (as e.g.~in \citet{AntonioPlat2014, Wuthrich2018, gabrielli2021individual}) this moving window evaluation enables a more thorough assessment of the sensitivity and general applicability of the model. On each evaluation date $\tau$ we train the models on the observed data (January, 2011 until $\tau$) and compare the out-of-sample reserve estimate over the next two calendar years with the actual claim development over the next two calendar years.

\subsubsection{Hierarchical reserving models} \label{sec:casestudy:calibration}

\paragraph{Hierarchical GLM} The hierarchical GLM follows the three layer structure \texttt{settlement}, \texttt{payment} and \texttt{size} defined in assumption (A3) and models each of these layers with a Generalized Linear Model (GLM). The settlement indicator is modelled using a binomial regression model and complementary log-log link function. We use a logistic regression model for the payment indicator and a gamma regression with log link function for the payment sizes. Actuaries are familiar with GLMs, given the long tradition of using GLMs in insurance pricing and reserving. Therefore, GLMs are the most likely candidate for supporting the transition from aggregate to individual reserving in practice. As is common in insurance pricing, we bin the continuous variables \texttt{age.insured}, \texttt{construction.year}, \texttt{property.value} and \texttt{rep.delay}. Table~\ref{table:covariate_bins} shows the chosen bins for each covariate. 

\begin{table}[ht!]
\centering
\begin{tabular}{l l}
\toprule
Variable & Bins \\ \midrule
\texttt{age.insured} & $[0, 39]$, $[40, 49]$ , $[50, 64]$, $65+$, NA \\
\texttt{construction.year} & $1950-$, $[1950, 1969]$, $[1970, 1984]$, $1985+$, NA \\
\texttt{property.value} & $\num{150000}-$, $(\num{150000}, \num{200000}]$, $(\num{200000}, \num{250000}]$, $\num{250000}+$, NA \\
\texttt{rep.delay} &  $5-$, $[5, 21]$, $21+$ \\
\bottomrule
\end{tabular}
\caption{List of chosen bins for the continuous covariates in the hierarchical GLM.}
\label{table:covariate_bins}
\end{table}

\begin{figure}[ht!]
\centering
\begin{subfigure}[t]{\textwidth}
\includegraphics[width = \textwidth]{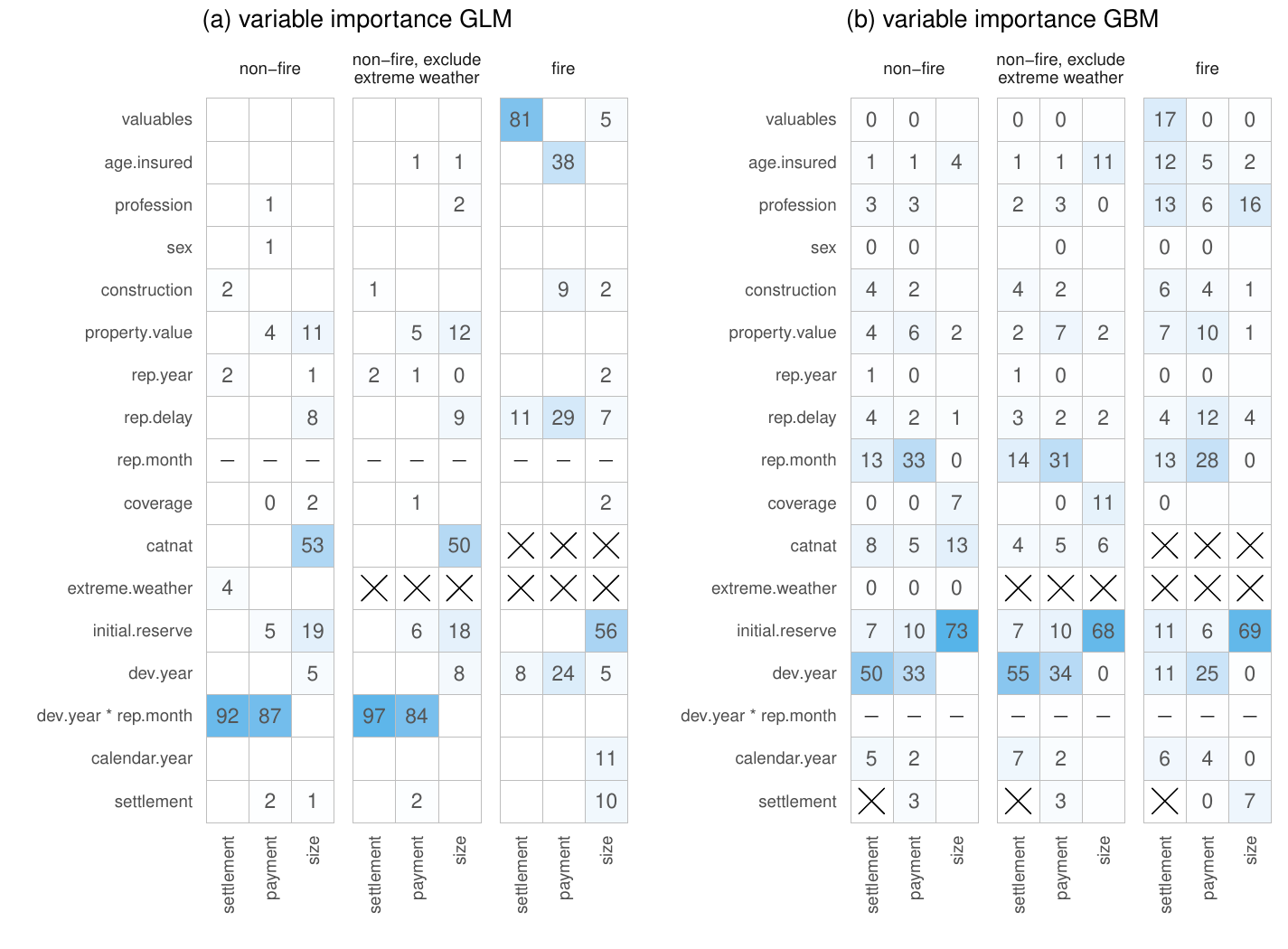}
\phantomcaption \label{figure:variable_importance_glm}
\phantomcaption \label{figure:variable_importance_gbm}
\end{subfigure}
\caption{Relative importance of the selected covariates in {\bf(a)} the hierarchical GLM and {\bf(b)} the hierarchical GBM.}
\label{figure:variable_importance}
\end{figure}

On the first evaluation date, January 1, 2015, we select the optimal set of covariates for each of the three GLMs (\texttt{settlement}, \texttt{payment} and \texttt{size}) using a forward selection procedure with $5$-fold cross-validation. In this approach, we start from the GLM with only an intercept. We then iteratively add the covariate that results in the largest increase in the weighted likelihood \eqref{eq:likelihood_general_weighted}, summed over all hold-out folds. We stop the iterative procedure when there are no more covariates that lead to an improvement in the weighted likelihood. In the moving window evaluation, we do not reselect the covariates on the other 364 evaluation dates, but recalibrate the parameters on each evaluation date using the most recent data.
Figure~\ref{figure:variable_importance_glm} shows the selected covariates in each GLM as well as a measure of the importance of each selected covariate. We compute the covariate's importance as the increase in the weighted likelihood \eqref{eq:likelihood_general_weighted}, summed over all hold-out folds, when sequentially adding covariates using the discussed forward selection procedure. These increases are rescaled per GLM and sum to 100. For non-fire claims the set of selected covariates changes only slightly when we omit extreme weather events. This is in line with the low importance assigned to the covariate \texttt{extreme.weather} when these claims are included. The interaction $\texttt{dev.year} * \texttt{rep.month}$ allows a more accurate determination of a claim's age, while using a yearly time grid in the reserving model. This is by far the most important determinant for the settlement and payment process of non-fire claims. 
Surprisingly, \texttt{dev.year} and $\texttt{dev.year} * \texttt{rep.month}$ have little effect on the size of non-fire claims. The most important determinants for the payment size are the claim type as coded in \texttt{catnat} and the \texttt{initial.reserve} set by the expert. The data set contains less fire claims and as a result fewer covariates are selected in the corresponding GLMs. Although these GLMs might be less predictive, the few selected covariates obtain high importance scores, since scores are rescaled to 100. This is in particular the case for \texttt{valuables} in layer \texttt{settlement} which receives an importance of 81 in the layer \texttt{settlement}, while having only a minor effect on the settlement probability. 

\paragraph{Hierarchical GBM} The hierarchical GBM follows the same three layer structure and the same distributional assumptions as in the hierarchical GLM, but models each layer with a tree based Gradient Boosting Machine (GBM). GBMs, as introduced by \cite{friedman2001}, model the data with a sequence of shallow decision trees, in which each tree improves the fit obtained with previous trees. The GBM has three major advantages. First, through a sequence of trees a non-linear effect can be estimated for continuous covariates, thus removing the need to bin continuous variables. Second, automatic feature selection is integrated in the calibration process. Third, simple interaction effects between the covariates are automatically modelled. As a result of these advantages, the covariates \texttt{age.insured}, \texttt{construction.year}, \texttt{property.value}, \texttt{rep.delay} and \texttt{initial.reserve} can be included as continuous covariates. Furthermore, we do not include the interaction $\texttt{dev.year} * \texttt{rep.month}$ as the model will automatically construct the relevant interactions. In return, a number of tuning parameters such as the number of trees and the depth of each tree have to be tuned. We tune the number of trees, the shrinkage and the interaction depth of each tree in the GBM on January 1, 2015 using the cross-validation strategy outlined in Section~\ref{section:covariateselection}. Once tuned, these parameters remain fixed throughout the 364 remaining evaluation dates. 

Figure~\ref{figure:variable_importance_gbm} shows the relative importance of the covariates in the various GBMs. The importance of a specific covariate is expressed as the sum of the improvements in the loss function each time that this covariate is used to split a tree in the GBM. We calculate this total improvement for each of the 365 evaluation dates. Then we average the results and scale it to 100. Since there is no explicit variable selection, importance is distributed over all covariates, which complicates the interpretation. For recent claims, \texttt{rep.month} allows for a more granular expression of the time elapsed since reporting of the claim, which is important when modelling the layer outcome variables \texttt{settlement}, \texttt{payment} and \texttt{size}. \texttt{initial.reserve} is the most important covariate when predicting the size of payments. This shows that claim experts base their reserve estimate on information of the claim beyond the covariates available in the modelling of the reserves. Similarly, the importance of other covariates shows that the practice of determining an initial reserve can be further improved by using a statistical model. The claim type \texttt{catnat}, which was important in the hierarchical GLM, becomes less important in the GBM.

\paragraph{Chain ladder method} We compare the individual hierarchical reserving models with the chain ladder method based on yearly aggregated data. As indicated in Section~\ref{section:aggregate_1}, the chain ladder method can be rephrased as a hiearchical reserving model with a single layer, i.e.~the payment size. On each evaluation date, we compute the RBNS reserve by applying the chain ladder method to a run-off triangle of payment sizes aggregated by reporting and development year. The choice for aggregating by reporting year results in an estimate for the RBNS reserve, as motivated in Section~\ref{section:aggregate}. Confidence bounds for the reserve estimate are derived from a normal assumption combined with the standard error under the Mack model \citep{Mack1999}.

\subsubsection{Evaluation of the RBNS reserve} \label{section:casestudy_rbns}
\begin{figure}[ht!]
\begin{subfigure}[t]{.9\textwidth}
\centering
\includegraphics[width = \textwidth]{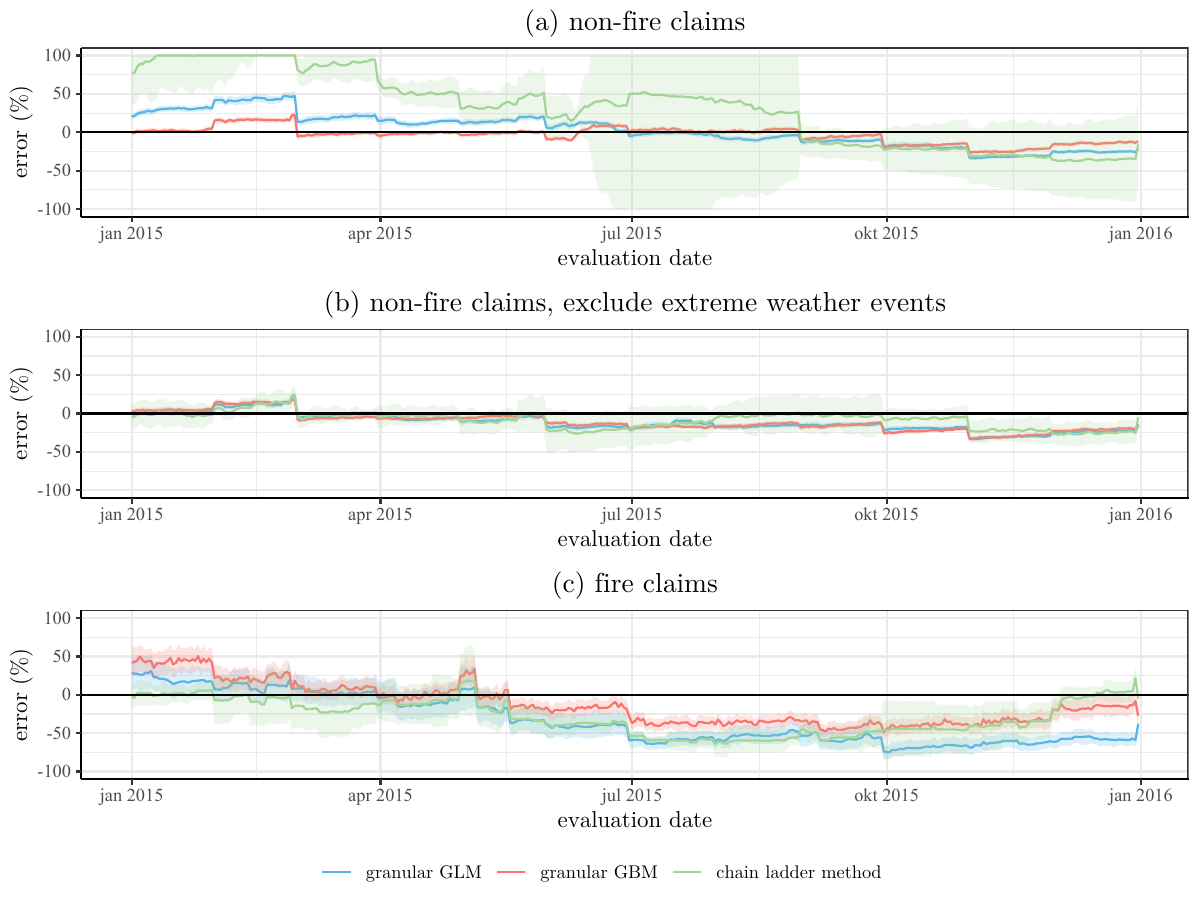}
\phantomcaption \label{figure:moving_window_rbns_other}
\phantomcaption \label{figure:moving_window_rbns_nonglobal}
\phantomcaption \label{figure:moving_window_rbns_fire}
\end{subfigure}
\caption{Percentage error when predicting the RBNS reserve on evaluation dates between January 1, 2015 and December 31, 2015 under the hierarchical GLM, hierarchical GBM and the chain ladder method. Errors are capped at $100\%$. {\bf(a)} shows the reserve for non-fire claims, {\bf (b)} the reserve for non-fire claims, when extreme weather events are excluded and {\bf (c)} the reserve for fire claims.}
\label{figure:moving_window_rbns}
\end{figure}

On each evaluation date we predict the expected RBNS reserve for the open claims over the next two years. We measure model performance via the percentage error of the predicted RBNS reserve compared to the actual RBNS reserve over the next two years, that is
\begin{align} \label{eq:PE.error}
\text{percentage error} = \frac{\texttt{predicted} - \texttt{actual}}{\texttt{actual}} \cdot 100\%.
\end{align} 
Figure~\ref{figure:moving_window_rbns} shows the evolution of the percentage error between January 2015 and December 2015 as obtained with the two hierarchical reserving models and the chain ladder method. The percentage error is capped at $100\%$ for improved readability of the figures. Table~\ref{table:rbns_performance} summarizes the daily errors by calculating the average percentage error and the average absolute percentage error over the 365 evaluation dates.

The reserve for non-fire claims (Figure~\ref{figure:moving_window_rbns_other}) combines the outstanding amounts on many small claims, which provides a rich data set for training reserving models. In a static setting without extreme events this would result in all models accurately predicting the reserve (see Figure~\ref{figure:moving_window_rbns_nonglobal} and Table~\ref{table:rbns_performance}). However, when extreme events enter the data, they produce outliers in the cells of the run-off triangle. This has a large impact on the chain ladder method, which fails to provide reasonable reserve estimates. This is a well known weakness of the chain ladder method and it is interesting that the individual models perform better, since they scale the reserve estimate automatically with the number of reported claims. In the granular GBM the performance was even slightly better when extreme events were included. The performance of the granular GLM, however, decreased slightly as this model can less flexibly distinguish both regimes. We observe higher prediction errors for all three models when predicting the reserve for fire claims (Figure~\ref{figure:moving_window_rbns_fire}). The combination of a low claim frequency and potentially high costs makes the reserve for fire claims difficult to predict. 

\begin{table}[ht!]
\centering
\begin{tabular}{ l  l   l  l l l l} \toprule 
\multirow{2}{*}{Portfolio} &  \multicolumn{2}{c}{hierarchical GLM} & \multicolumn{2}{c}{hierarchical GBM} & \multicolumn{2}{c}{chain ladder} \\   
& $\mu(PE)$ & $\mu(|PE|)$ & $\mu(PE)$ & $\mu(|PE|)$ & $\mu(PE)$ & $\mu(|PE|)$  \\ \midrule
non-fire claims  & { 2.50} & {18.67} & -3.61 & {\bf 8.08} & 33.89 & 51.31\\  
\rule{0pt}{14pt}non-fire claims,  & \multirow{2}{*}{-11.54} & \multirow{2}{*}{{13.92}} & \multirow{2}{*}{{-11.05}} & \multirow{2}{*}{14.12} & \multirow{2}{*}{-8.21} & \multirow{2}{*}{{\bf 9.97}} \\ 
exclude extreme weather \\ 
\rule{0pt}{14pt}fire-claims   &  -33.17 & {39.09} & {-12.69} & {\bf 25.92} & -28.41 & 29.76\\\bottomrule
\end{tabular}
\caption{Evaluation of the average performance of the hierarchical GLM, hierarchical GBM and chain ladder method over 365 evaluation dates between January 1, 2015 and December 31, 2015. Average performance is expressed as the mean percentage error and the mean absolute percentage error.}
\label{table:rbns_performance}
\end{table}

\section{Scenario testing with simulated portfolios}\label{section:scenariotesting}
We evaluate the predictive performance of our hierarchical reserving model on the portfolios generated by a simulation machine. Inspired by \cite{gabrielli2018individual} and \cite{avanzi2021synthetic} we design a simulation machine to generate portfolios registering the development of individual claims over time, along several scenarios introduced in Section~\ref{sec:intro.scenarios}. This simulation engine generates the development of individual claims in continuous time, with the occurrence of a claim, its reporting, payments and settlement as the recorded events. These events are influenced by claim characteristics, some observable and some unobservable by the analyst. Appendix~\ref{appendix:sim.machine} provides a detailed description of the technical set-up of this simulation machine for the base scenario. Figure~\ref{tikz:structuredata} displays the structure of a simulated dataset. The insurer's interest is to predict the total RBNS reserve for claims reported between January 1, 2012 ($t_{\min}$) and December 31, 2020 ($t_{\max}$), leading to a 9-year observation window. We set the evaluation date to $\tau =$ December 31, 2020 and track every claim over a maximum of 9 years after its reporting date. The simulation machine is further equipped with two 3-level categorical characteristics, one that represents the \texttt{type} of a claim  (levels \texttt{T1}, \texttt{T2} and \texttt{T3}) and one unobservable covariate \texttt{hidden} (levels \texttt{L}, \texttt{M}, \texttt{H}). The insurer is only aware of the type of the claim. The four scenarios outlined below focus on realistic events that impact the development of claims over time and hence the RBNS reserves.

\begin{figure}[ht!]
\centering
\begin{tikzpicture}[snake=zigzag, line before snake = 2mm, line after snake = 2mm, text width = 3cm, align = left]
    \draw[->] (0,0) -- (5.2,0);
    \draw[->] (0,0) -- (0,-5.2);
    \draw[dashed] (0,-4.5) -- (4.5,0);
    \draw (0,-4.5) -- (4.5,-4.5);
    \draw (4.5,-4.5) -- (4.5,0);
    
    \fill[gray!30]  (0.025,-0.025) -- (4.475,-0.025) -- (0.025,-4.475) -- cycle;
    
  	\foreach \x in {0.25,0.50,0.75,1,1.25,1.50,1.75,2,2.25}
      \draw[gray!30] (\x,-4.5+\x) -- (2*\x,-4.5);
     \foreach \x in {2.5,2.75,3,3.25,3.5,3.75,4,4.25,4.5}
      \draw[gray!30] (\x,-4.5+\x) -- (4.5,-4.5*2 + 2*\x);

    \draw (0,0)   node[left, align = right] {\small January 1, 2012};
    \draw (0,-4.5) node[left, align = right] {\small December 31, 2020};
    \draw (0,-0.5)   node[left, align = right] {\scriptsize Claim 1};
    \draw (0,-1.3)   node[left, align = right] {\scriptsize Claim 2};
    \draw (0,-2.8)   node[left, align = right] {\scriptsize Claim 3};
    \draw (0,-4.0)   node[left, align = right] {\scriptsize Claim 4};
    \draw (0,-4.9) node[left, align = right] {\small (evaluation date)};
    \draw (4.5,0) node[align = center, above = 1pt] {\small 9 years};
    \draw (5.2,0) node[right] {\small Settlement delay};
    \draw (0,-5.2) node[align = center, below = 3pt] {\small Reporting date};
 
   \draw (0,0) node[circle,fill,inner sep=0.01pt, text width = 0.1cm]{};
   \draw (4.5,0) node[circle,fill,inner sep=0.01pt, text width = 0.1cm]{};
   \draw (0,-4.5) node[circle,fill,inner sep=0.01pt, text width = 0.1cm]{};
   
   \foreach \x in {-0.5,-1.3,-2.8,-4}
      \draw (0,\x) node[draw,circle,fill=white,inner sep=0.5pt,text width=0.15cm]{};
    
    \draw (0.1,-0.5) -- (3.2,-0.5);
    \draw (0.1,-1.3) -- (3.9,-1.3);
    \draw (0.1,-2.8) -- (1.05,-2.8);
    \draw (0.1,-4.0) -- (2.6,-4.0);
    
\draw (3.2,-0.5) node[thick,draw,cross out,inner sep=0pt,minimum width=2.5pt,minimum height=5pt,text width = 0.2cm] {};  
\draw (3.9,-1.3) node[thick,draw,cross out,inner sep=0pt,minimum width=2.5pt,minimum height=5pt,text width = 0.2cm] {}; 
\draw (1.05,-2.8) node[thick,draw,cross out,inner sep=0pt,minimum width=2.5pt,minimum height=5pt,text width = 0.2cm] {}; 
\draw (2.6,-4.0) node[thick,draw,cross out,inner sep=0pt,minimum width=2.5pt,minimum height=5pt,text width = 0.2cm] {}; 
\end{tikzpicture}
\caption{Structure of a simulated dataset. We simulate both the gray and the hatched triangle. The insurer's interest is to predict the RBNS reserve for claims that are reported between January 1, 2012 and December 31, 2020. The gray triangle shows the data observed on the evaluation date December 31, 2020 and will be used to train the models and to predict the hatched triangle. The white dots and the crosses represent the reporting date and the settlement date respectively. The developments of claims 1 and 3 are fully observed at the evaluation date, whereas the developments of claims 2 and 4 are right censored at the evaluation date. \label{tikz:structuredata}}
\end{figure}

\subsection{Data characteristics}\label{sec:datachac}
To align with the discrete time hierarchical reserving model discussed in this paper, we discretize the continuous time portfolios generated by the simulation engine to annual portfolios. The resulting annual portfolio contains up to 9 records per reported claim, one for each development year since reporting. Each record of a claim lists the annual payment size in the corresponding development year. The total of these annual payment sizes over the 9 development years associated to one specific claim is referred to as the claim size. In addition, we define a payment indicator that indicates if a payment takes place within a particular development year. To construct and calibrate the hierarchical reserving models, we only use the discretized information and the reporting month of the claim. We do not use the continuous time covariates. Appendix~\ref{appendix:sim.machine} provides more information and shows the records of an exemplary claim.

\begin{figure}[ht!]
\includegraphics[width = \textwidth]{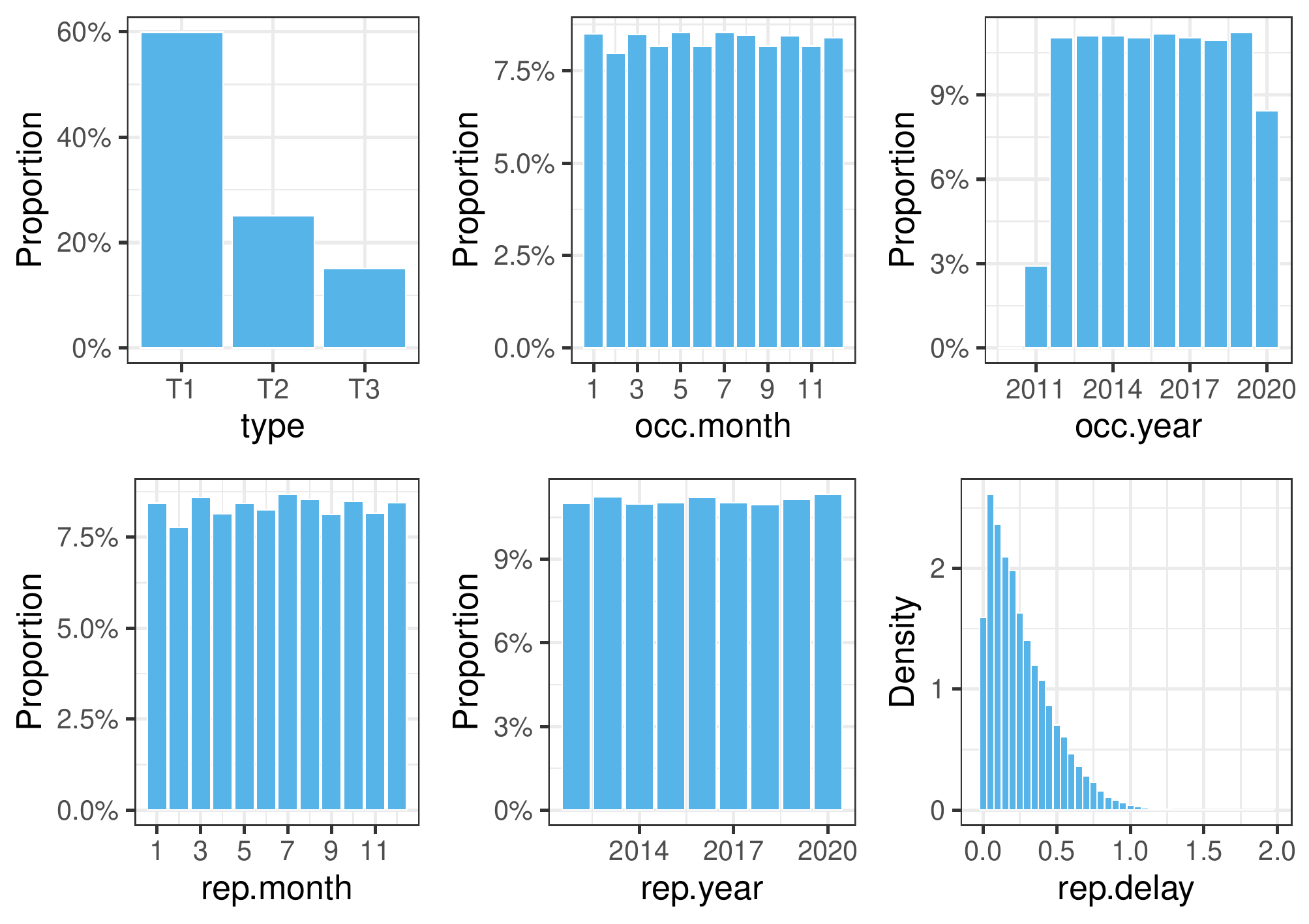}
\caption{Distribution of the claim covariates in one portfolio simulated along the baseline scenario. We generate $125\ 000$ claims and use the data that is available at the evaluation date December 31, 2020 (gray triangle in Figure~\ref{tikz:structuredata}). \label{fig:sc1:claimcov}}
\end{figure}

\begin{figure}[ht!]
\includegraphics[width = \textwidth]{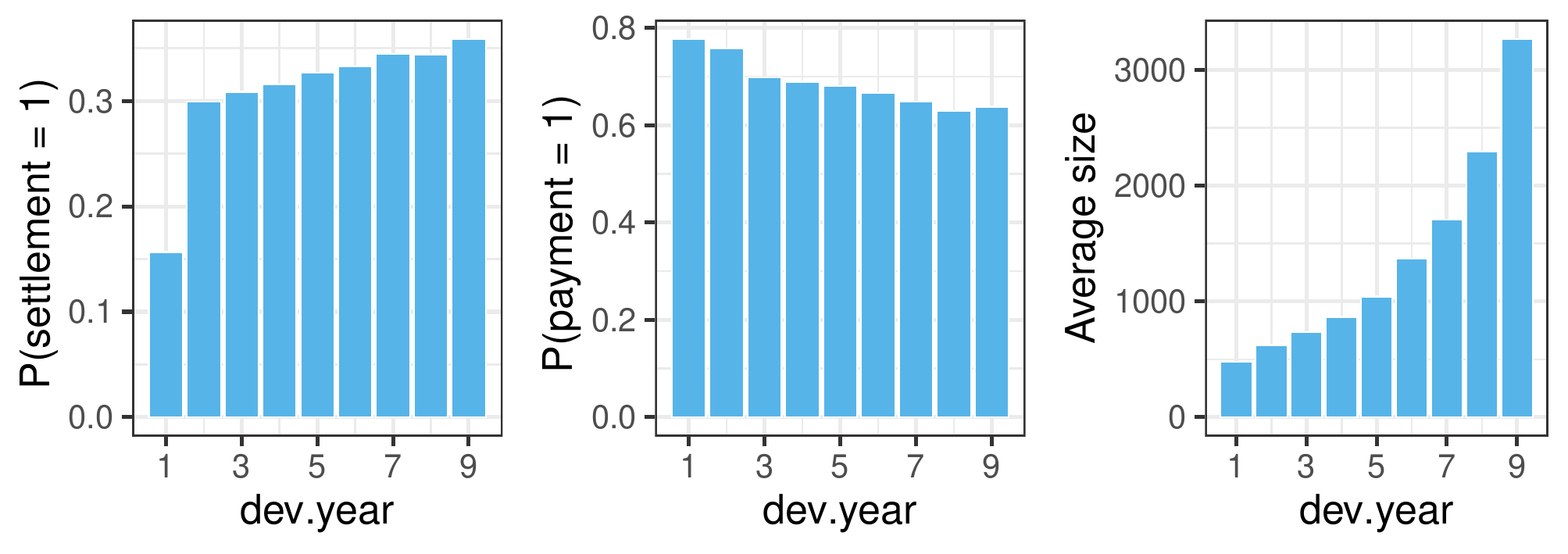}
\caption{The empirical probability that a claim settles given that the claim is still open (left), the empirical probability of a payment given that the claim is still open (middle) and the average payment size given that a payment takes place (right) in one portfolio simulated along the baseline scenario with data available up to the evaluation date December 31, 2020. We show the results per development year since reporting. \label{fig:sc1:layercov}}
\end{figure}

Table~\ref{table:Sim.Machine:covariates} lists the covariates that are generated with the simulation engine. Figure~\ref{fig:sc1:claimcov} shows barplots and histograms of some of the claim covariates for a portfolio generated along the baseline scenario. In this example, about $60\%$ of the reported claims are of Type 1, $25\%$ of Type 2 and $15\%$ of Type 3. These percentages as well as the distributions of the other covariates correspond to the sampling distributions specified upfront in our simulation engine. Claims with an occurrence year of 2010 or 2011 may be present in the generated portfolio if their reporting takes place after January 1, 2012. Figure~\ref{fig:sc1:layercov} displays the distribution of the layer outcome variables as a function of the development year since reporting. For open claims, the settlement probability increases in later development years, whereas the likelihood of registering a payment decreases over time. However, conditional on making a payment, the expected payment size is larger in later development years. The settlement probability is lower in the first development year, since this development year is shorter than later development years as it only runs from the reporting date of a claim until December 31 of that same year.

\begin{table}[ht!]
\centering
\begin{tabularx}{\textwidth}{>{\hsize=.2\hsize}X >{\hsize=.1\hsize}X >{\hsize=.7\hsize}X}
\toprule
\multicolumn{3}{l}{{\bf Claim covariates}}  \\
\texttt{occ.date} & Character &  Date on which the accident occurred: `yyyy-mm-dd'. \\
\texttt{occ.year} & Factor & Year in which the accident occurred: $2010$, $2011$, $\ldots$, $2020$. \\
\texttt{occ.month} & Factor & Month of the year in which the accident occurred: $1$, $2$, $\ldots$, $12$. \\
\texttt{rep.date} & Character & Date on which the claim is reported to the insurer: \mbox{`yyyy-mm-dd'}. \\
\texttt{rep.year} & Factor & Year in which the claim was reported to the insurer: $2012$, $2013$, $\ldots$, $2020$. \\
\texttt{rep.month} & Factor & Month of the year in which the claim was reported to the insurer: $1$, $2$, $\ldots$, $12$. \\
\texttt{rep.delay} & Numeric & Delay in years between the occurrence and reporting of the claim. \\
\texttt{settlement.date} & Character & Date on which the claim settles: \mbox{`yyyy-mm-dd'}. \\
\texttt{settlement.year} & Factor & Year in which the claim settles. \\
\texttt{type} & Factor & Type of the claim: $\texttt{T1}$, $\texttt{T2}$, $\texttt{T3}$. \\
\texttt{hidden} & Factor & Covariate that is assumed to be hidden from the insurer and which cannot be used to predict the total RBNS reserves: $\texttt{L}$, $\texttt{M}$, $\texttt{H}$. \\
\multicolumn{3}{l}{{\bf Development covariates}} \\ 
\texttt{dev.year} & Factor & The number of years elapsed since the claim is reported to the insurer: $1$, $2$, $3$, $\ldots$, $9$. Development year $1$ corresponds to the year in which the claim is reported to the insurer.\\
\texttt{calendar.year} & Numeric & Number of years elapsed between the start of the portfolio (year 2012) and \texttt{dev.year}: 1, 2, 3, $\ldots$ \\
\multicolumn{3}{l}{{\bf Layer outcome variables}} \\ 
\texttt{settlement} & Binary & Settlement indicator. The claim settles in the current development year: yes (1) or no (0). \\
\texttt{payment} & Binary & Payment indicator. A payment occurs in the current development year: yes (1) or no (0). \\
\texttt{size} & Numeric & Total amount paid in the current development year. \\
\bottomrule
\end{tabularx}
\caption{List of covariates and response variables available in a portfolio generated with the simulation machine outlined in Appendix~\ref{appendix:sim.machine}.}
\label{table:Sim.Machine:covariates}
\end{table}

\begin{figure}[ht!]
\includegraphics[width = 0.9\textwidth]{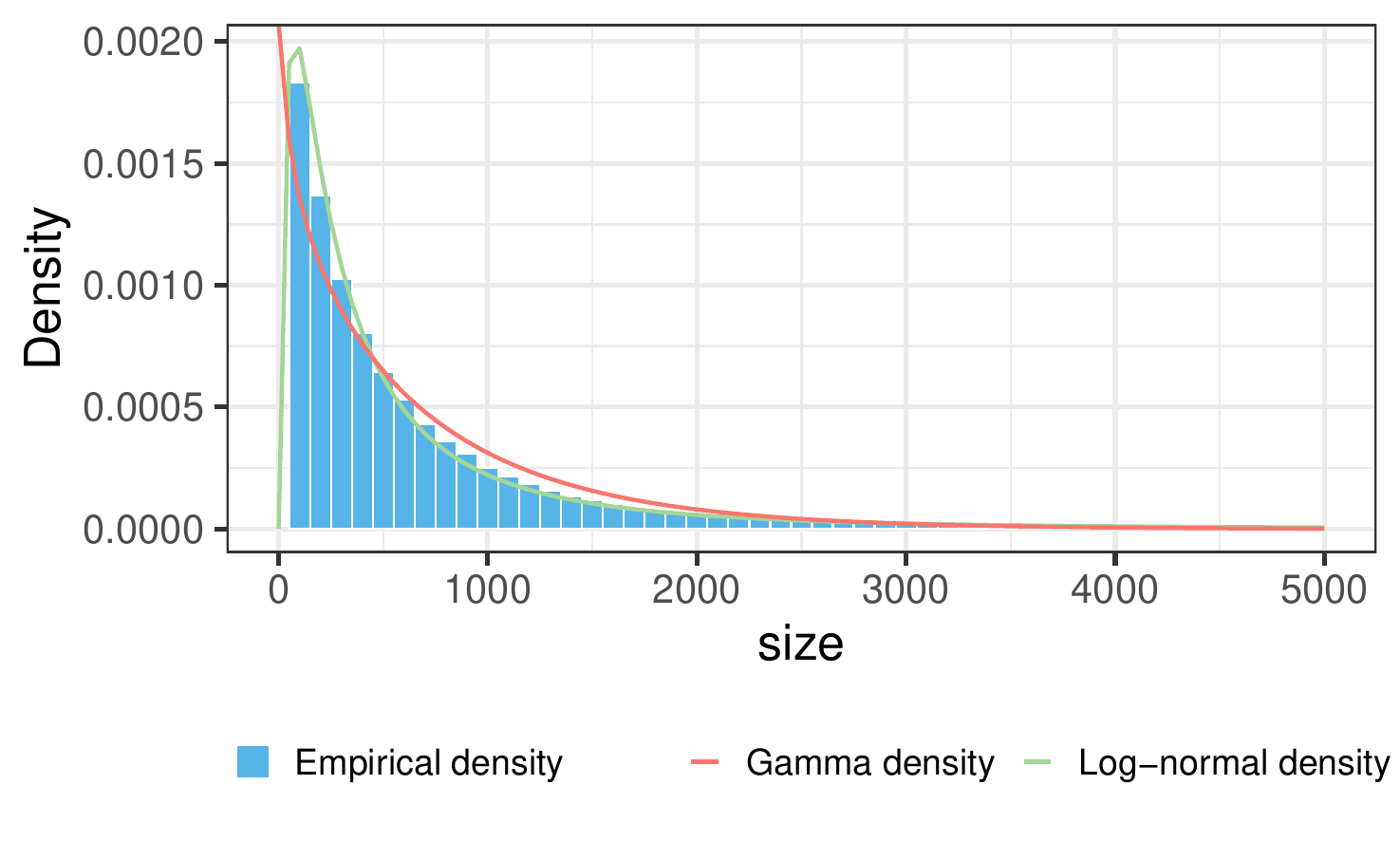}
\caption{Empirical distribution of the yearly, aggregated payment sizes, given that a payment occurred in blue. The red and green line show the estimated gamma and log-normal density function respectively. Estimation is done using maximum likelihood and on one portfolio simulated along the baseline scenario with data available up to the evaluation date December 31, 2020. \label{fig:sizedistr}}
\end{figure}

Figure~\ref{fig:sizedistr} shows the empirical density function of the yearly, aggregated payment sizes, given that a payment occurs, in one generated portfolio. Green and red lines show the fitted gamma and log-normal density function respectively, using maximum likelihood estimation. The log-normal density fits the data better than the gamma density. This is not surprising, as the payment sizes (in the continuous time setting) where generated from a log-normal distribution. However, in Section~\ref{section:calibrationhrm} we use the gamma distribution to model the payment size layer. This is a common assumption in modelling severities in insurance pricing since predictions obtained with a log-normal regression model are prone to transformation bias.

\subsection{Introducing the scenarios} \label{sec:intro.scenarios}

\paragraph{Scenario 1: Baseline scenario.} In the baseline scenario, we assume that $60\%$ of the claims in the portfolio are of Type 1, $25\%$ of Type 2 and $15\%$ of Type 3. We further assume that the time between two consecutive, non-aggregated payments in the continuous time setting are exponentially distributed with \texttt{type}-dependent parameters. The corresponding payment sizes (in the continuous time setting) are log-normally distributed with a fixed variance, but with an increasing mean for payments occurring in later development years since reporting. The settlement delay, i.e.~the time between the reporting and the settlement date, is assumed to be beta distributed again with \texttt{type}-dependent parameters. We provide the full simulation strategy for generating a portfolio along the baseline scenario in Appendix~\ref{appendix:sim.machine}. All other scenarios start from this baseline scenario, but are subject to minor changes in either the distribution of the claim type in the portfolio (scenario 2), the claim occurring process (scenario 3) or the claim settlement process (scenario 4). 

\paragraph{Scenario 2: Change in claim mix.} In this scenario, we assume that the distribution of the type of the claims in the portfolio changes with each subsequent occurrence year. We start from the baseline scenario in the first occurrence year $2010$. In contrast to the baseline scenario, we assume in each consecutive year a decrease in the number of Type 1 claims by $2\%$, an increase of Type 2 claims by $0.5\%$ and an increase of Type 3 claims by $1.5\%$. Let us denote the occurrence date random variable by $A$. Then we apply the following adaptation to the distribution of the claim types in the portfolio:
\begin{align*}
\mathbb{P}(\texttt{type} = \tau \mid \text{year}(A) = s) = p_{\tau} + \nu_{\tau} (s - 2010), \hspace{1cm} & \tau \in \{\texttt{T1}, \texttt{T2}, \texttt{T3}\} 
\end{align*}
where $(p_{\texttt{T1}}, p_{\texttt{T2}}, p_{\texttt{T3}}) = (0.60,0.25,0.15)$, $(\nu_{\texttt{T1}}, \nu_{\texttt{T2}}, \nu_{\texttt{T3}}) = (-0.02,0.005,0.015)$ and $\text{year}(A) = s$ indicates that the claim occurred in year $s$.

\begin{figure}[ht!]
\centering
\includegraphics[width = \textwidth]{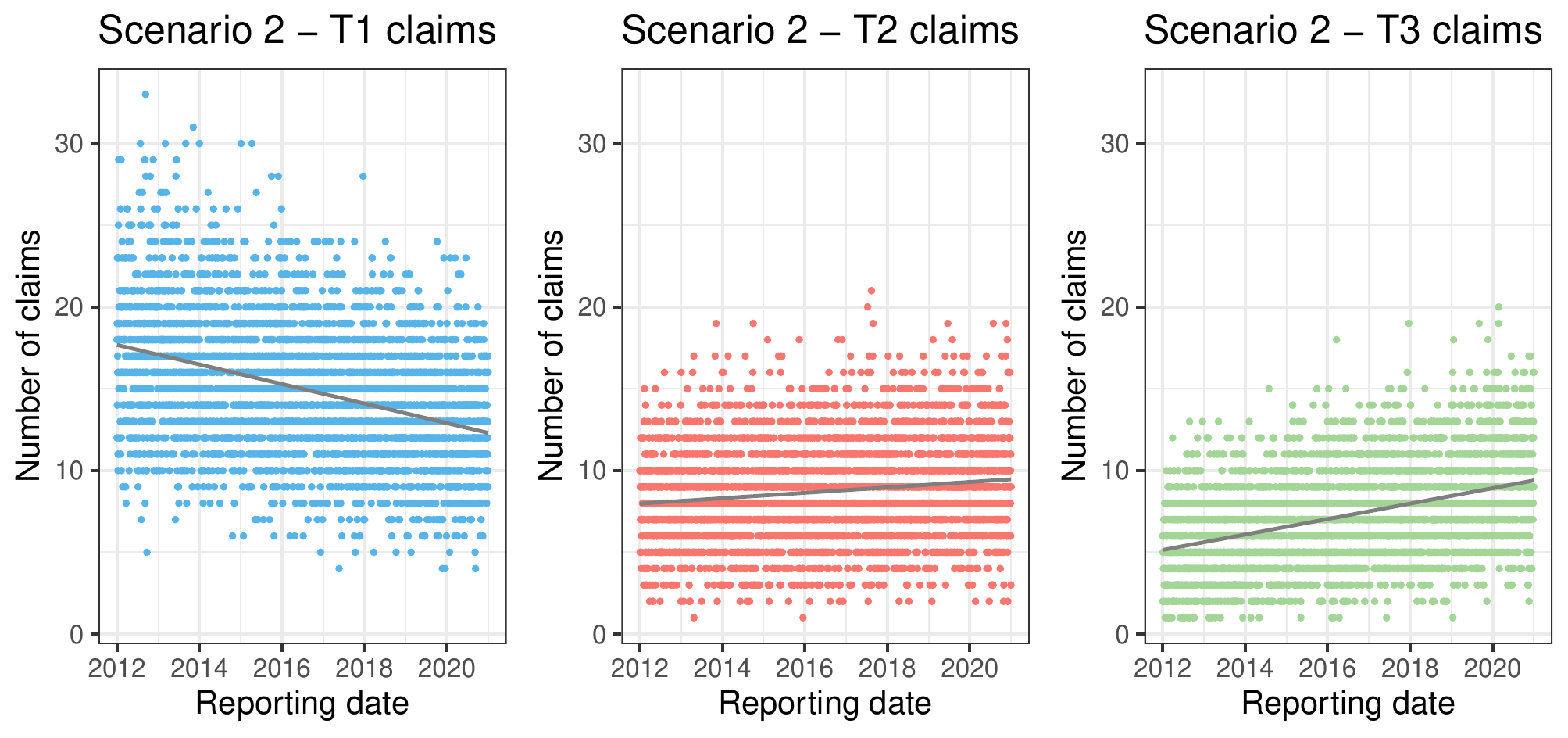}
\caption{The coloured dots represent the number of claims at each reporting date between $t_{\min}$ (January 1, 2012) and $t_{\max}$ (December 31, 2020). The gray solid line is a linear model fit. The results are based on a portfolio of $125\ 000$ claims simulated along scenario 2 with the change in claim mix.\label{fig:sc2:expolore}}
\end{figure}

Figure~\ref{fig:sc2:expolore} shows the number of simulated claims per claim type for each reporting date. We clearly observe the decrease in the number of Type 1 claims and the increase of Type 2 and 3 claims over time, in line with the implemented change in claim mix in scenario 2.

\paragraph{Scenario 3: Extreme event.} In this scenario more claims occur in a certain period as a result of an extreme event, for example a very cold winter could potentially cause more car accidents or a very hot summer could cause more fire damages. We simulate this by increasing the probability of generating an occurrence date in some extreme event period $[t_{\min}^e, t_{\max}^e]$. The start date of the extreme event $t_{\min}$ is sampled randomly in the year 2019 for each generated portfolio and the duration of the event is set to 30 days. On days within the extreme event period, five times more claims occur than on a regular day. We hence choose the following discrete distribution for the occurrence date r.v. $A$:
\begin{align*} \label{eq:scenario3}
\mathbb{P}[A = t] =  \begin{cases}
\dfrac{1}{n_T + 4\cdot n_T^e} & \text{ if } t \not \in [t_{\min}^{e}, t_{\max}^e] \\ \\
\dfrac{5}{n_T + 4\cdot n_T^e} & \text{ if } t \in [t_{\min}^e, t_{\max}^e], 
\end{cases}
\end{align*}
where $n_T$ is the total number of possible occurrence dates in the portfolio, namely the number of days between January 1, 2010 and December 31, 2020 (see Appendix~\ref{appendix:sim.machine}). In addition, $n_T^e := \text{days}(t_{\min}^e,t_{\max}^e)$ denotes the total number of days in the extreme event period $[t_{\min}^e, t_{\max}^e]$. This is equal to $30$ days by assumption. The probabilities are chosen such that they add up to one and such that the probability of generating an occurrence date belonging to the extreme event period $[t_{\min}^e, t_{\max}^e]$ is five times as large as generating an occurrence date that does not belong to the extreme event period.

\begin{figure}[ht!]
\centering
\includegraphics[width = \textwidth]{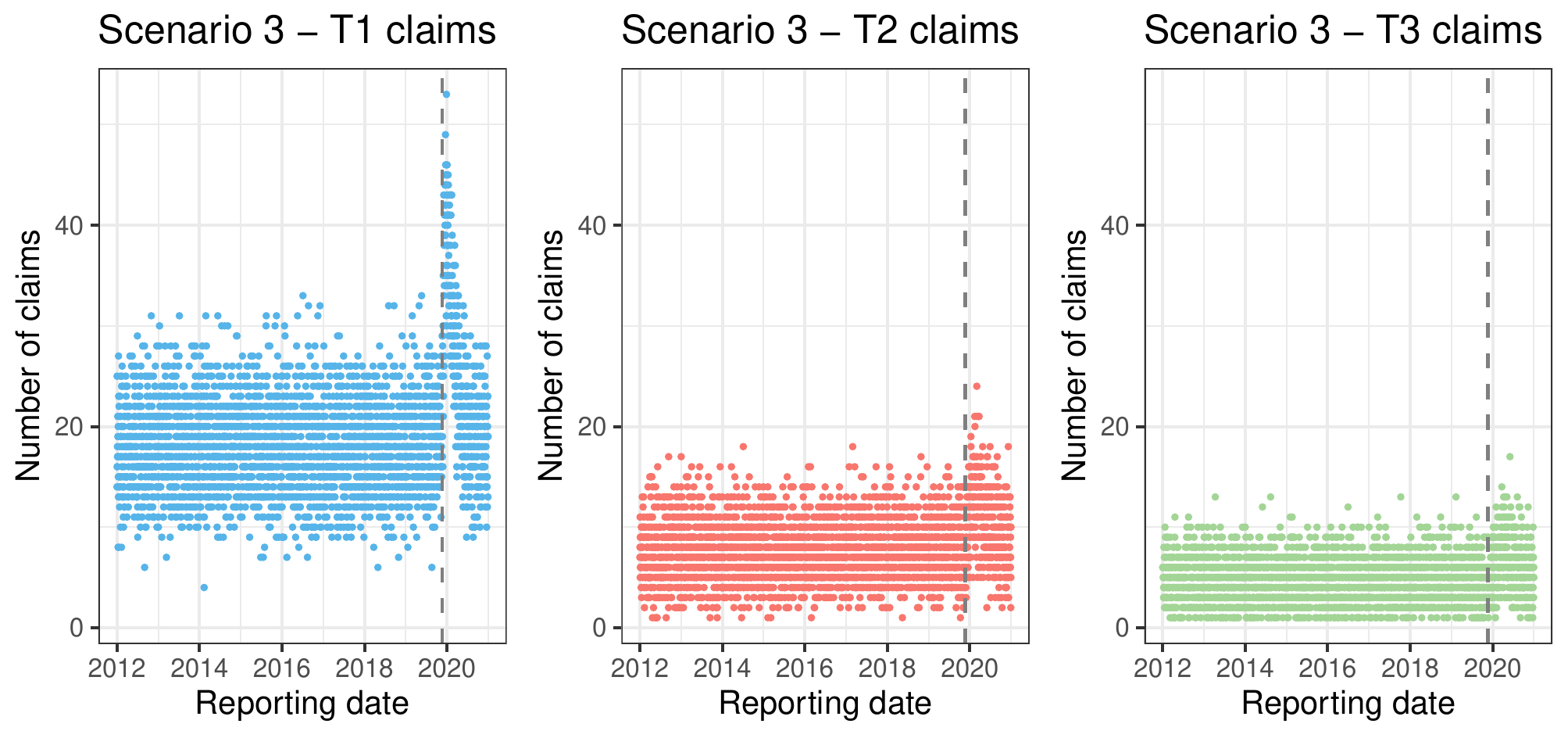}
\caption{The coloured dots represent the number of claims at each reporting date between $t_{\min}$ and $t_{\max}$. The gray dashed line indicates the start of the extreme event period, i.e.~November 20, 2019 for this particular portfolio. The results are based on a portfolio of $125\ 000$ claims simulated along the extreme event scenario (scenario 3).\label{fig:sc3:explore}}
\end{figure}

Figure~\ref{fig:sc3:explore} visualizes the number of reported claims per reporting date. The gray dashed line indicates the start date $t_{\min}^e$ of the extreme event period, namely November 20, 2019 for this specific portfolio. We observe a surplus of reported claims after this period because of the extreme event period. 

\paragraph{Scenario 4: Change in settlement delay.} This scenario assumes a drastic change in the claim settlement. In practice, this could be the result of a change in regulations or an improvement in the efficiency of the claims handling process. We impose a 15\% improvement in the claim settlement delay for claims occurring after January 1, 2017. More specifically, we multiply the settlement delay $d_k^s$ of a claim $k$, that occurred after January 1, 2017, by a factor of 0.85. In practice, a change in the settlement delay will most likely also entail a change in the payment delays of a claim. Therefore, we multiply the continuous time between two consecutive, non-aggregated payment dates by a factor of 0.85 for those claims that occur after January 1, 2017. In addition, we model the corresponding payment sizes (in continuous time setting) as if no reduction in payment delays occurred. In this way, the payments for claims occurring after January 1, 2017 are more concentrated but the payment sizes remain the same.

\begin{figure}[ht!]
\centering
\includegraphics[width = \textwidth]{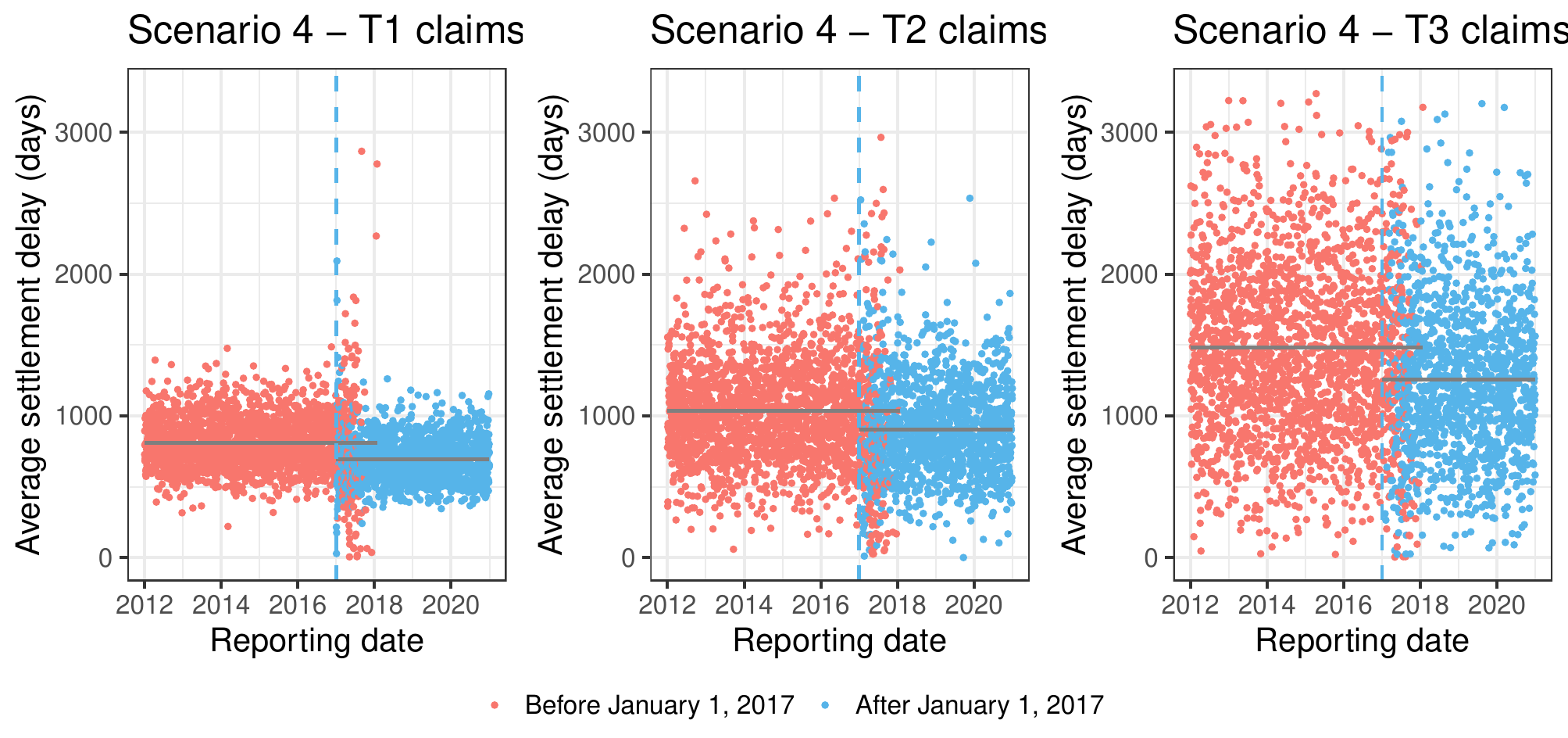}
\caption{The coloured dots represent the average settlement delay (in days) of claims reported at the same reporting date between $t_{\min}$ and $t_{\max}$. We make a distinction between claims reported before (red) and after (blue) January 1, 2017. The gray dashed lines denote the average settlement delay of all the claims reported before and after January 1, 2017 respectively. The vertical dashed line indicates the change point January 1, 2017. The results are based on a portfolio of $125\ 000$ claims simulated along the change in settlement delay scenario (scenario 4).\label{fig:sc4:explore}}
\end{figure}

We visualize a portfolio simulated along this scenario in Figure~\ref{fig:sc4:explore} by displaying the average settlement in days as a function of the reporting date. As in the baseline scenario, Type 1 claims have the lowest settlement delay whereas Type 3 claims have the largest settlement delay on average. However, as the gray dashed lines indicate, we now observe a decrease in the average settlement delay from 2017 onwards.

\begin{figure}[ht!]
\centering
\includegraphics[width = \textwidth]{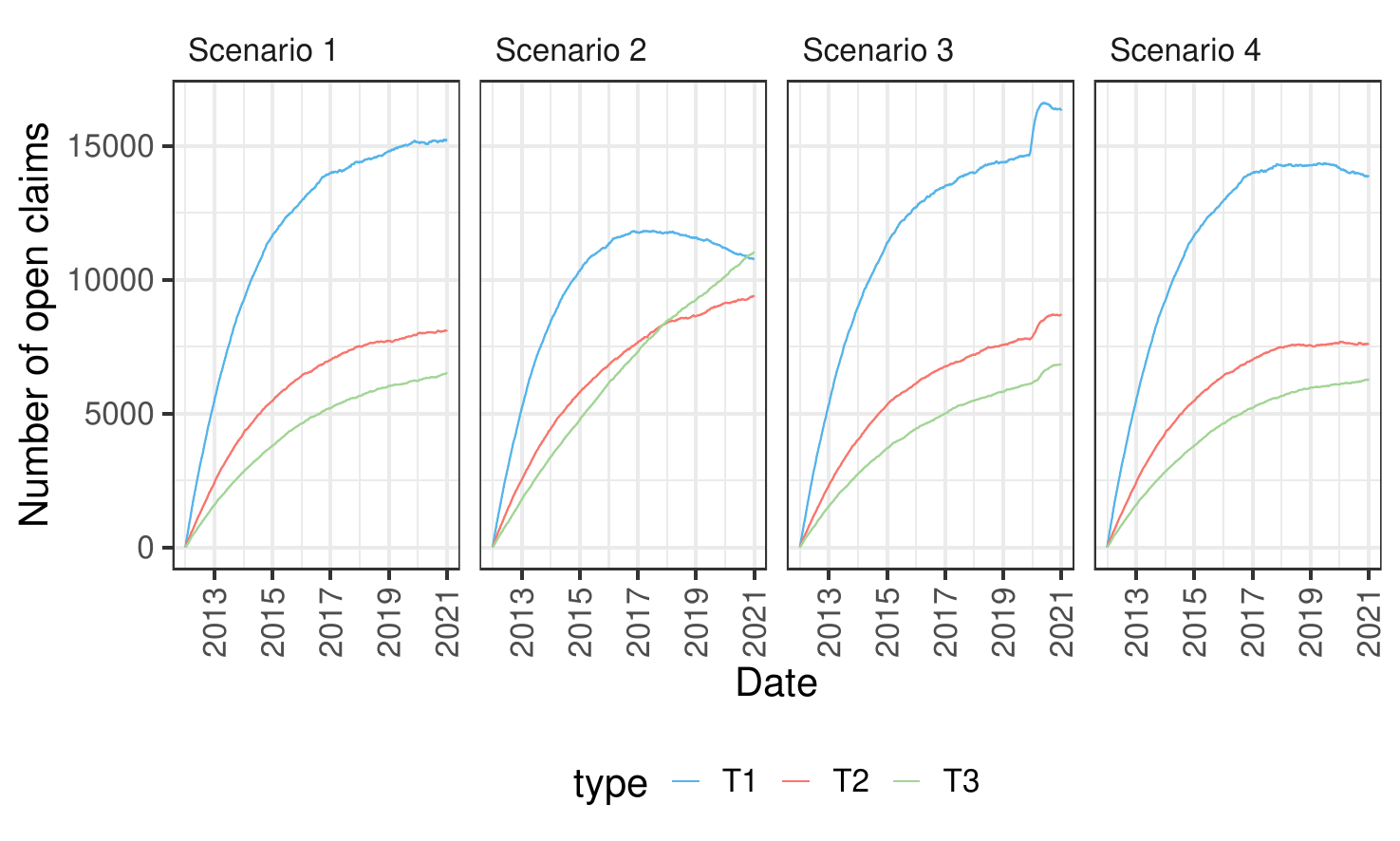}
\caption{The number of open claims at each date between $t_{\min}$ (i.e.~January 1, 2012) and $t_{\max}$ (i.e.~December 31, 2020). The results are based on a portfolio of $125\ 000$ claims simulated along each considered scenario. \label{fig:comparescenarios}}
\end{figure}

\paragraph{Comparing the four scenarios.} To conclude our exploratory analysis we compare one simulated portfolio from each considered scenario by visualizing the number of reported, open claims on any date between $t_{\min}$ and $t_{\max}$ in Figure~\ref{fig:comparescenarios}. In the claim mix change scenario (scenario 2), we clearly observe less claims of Type 1 due to a steady decrease in the number of claims each consecutive occurrence year. The linear increase in the number of claims of Type 2 and Type 3 each consecutive occurrence year is reflected in more open claims of these types over time. In the third panel, we observe a bump in the number of open claims (of any type) around the year 2020. This corresponds to the extreme event period starting at November 20, 2019. Due to the reporting delay of the claims, this is reflected some time later. In the last panel, we observe a flattening of the curves from the year 2017 onwards. Since we reduce the settlement delay of the claims by a factor 0.85, this is again in line with our expectations.

\subsection{Calibration of the hierarchical reserving models}\label{section:calibrationhrm}
We simulate 100 portfolios along each scenario using the proposed simulation engine. On each generated portfolio we calibrate three reserving models, namely the hierarchical GLM, the hierarchical GBM and the chain ladder method, on the evaluation date December 31, 2020. We calibrate the hierarchical reserving models as outlined in Section~\ref{section:covariateselection}.

\paragraph{Hierarchical GLM.} The hierarchical GLM consists of the three layer structure (\texttt{settlement}, \texttt{payment} and \texttt{size}) as defined in Assumption (A3). We use the same GLMs with the same link functions as in the case study (Section~\ref{sec:casestudy:calibration}) to model the three layers. For each scenario, we select the optimal covariate set for each layer on one simulated portfolio using a forward selection procedure with a 5-fold cross-validation. We iteratively add the covariate to our model that leads to the largest increase in the weighted likelihood in~\eqref{eq:likelihood_general_weighted} over all hold-out folds. For computational reasons we only apply this selection procedure on the first simulated portfolio from each scenario and reuse the selected covariates on all other simulated portfolios from that scenario.

\begin{figure}[ht!]
\centering
\includegraphics[width = \textwidth]{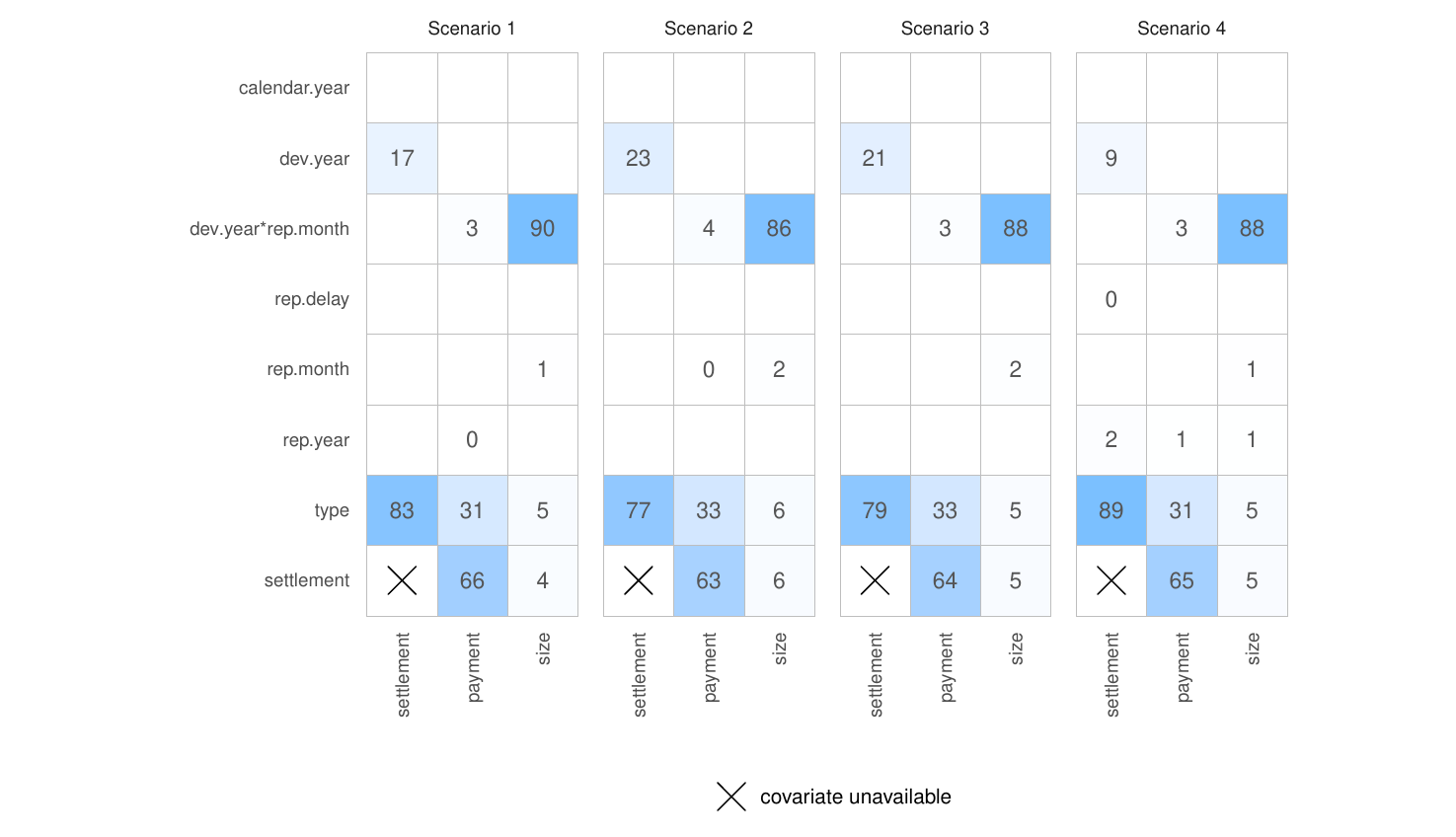}
\caption{The relative variable importance of the selected covariates in one simulated portfolio along each scenario in the hierarchical GLM. The cross-sign indicates that the covariate is not available, e.g.~the covariate \texttt{settlement} is obviously not available in the layer \texttt{settlement}. Numbers are rounded to the nearest integer. An empty cell indicates that the covariate was not selected during the covariate selection procedure. \label{fig:importance_glm} }
\end{figure}

The available covariates are listed in Table~\ref{table:Sim.Machine:covariates}. We only use the covariates that are represented in a discretized format to stay in line with the discrete time hierarchical reserving model discussed in this paper. In addition, we do not use covariates that provide information about the occurrence dates of the claims (\texttt{occ.year}, \texttt{occ.month}) as the reporting year and the reporting delay of the claims are considered as sufficient. Similar to the case study in Section~\ref{section:caseStudy}, we add the interaction effect $\texttt{dev.year} * \texttt{rep.month}$ to allow for a more accurate determination of the claim's age. Figure~\ref{fig:importance_glm} lists the variable importance of the selected covariates for each layer of the hierarchical GLM and for each scenario. The development year since reporting (\texttt{dev.year}) and the type of the claim (\texttt{type}) are the most important covariates in the layer \texttt{settlement}, across all four scenarios. This is because the expected claim settlement delay in the simulation machine is the lowest for claims of Type 1 and the highest for claims of Type 3. Although in scenario 4 the development process speeds up from 2017 onwards, the importance of the covariate \texttt{rep.year} is rather limited. In addition, the reporting year and reporting delay serve as a proxy for the occurrence year of the claim. Next, the most important covariates in the layer \texttt{payment} are the type of the claim and the settlement indicator (\texttt{settlement}) of the claim. The importance of the variable \texttt{type} is obvious since we simulate the continuous time between two consecutive payments from an exponentially distributed random variable with a type-dependent parameter. At last, the interaction effect $\texttt{dev.year} * \texttt{rep.month}$ clearly has the highest importance in the layer \texttt{size}. This is because the non-aggregated payment sizes (in the continuous time setting) are generated from a log-normal distribution where the mean depends on the payment delay. We only observe minor differences across the four scenarios.

\paragraph{Hierarchical GBM.} The hierarchical GBM uses the same three layer structure with \linebreak \texttt{settlement}, \texttt{payment} and \texttt{size}. We use the same distributional assumptions as in the hierarchical GLM, but fit each layer with a GBM instead of a GLM. Advantages of the GBM over the GLM are discussed in Section~\ref{section:caseStudy}. We tune some of the GBM parameters, namely the number of trees, the shrinkage and the interaction depth of each tree in line with the approach in Section~\ref{sec:casestudy:calibration}. Table~\ref{table:tunedcov} shows the tuned parameters for each layer and for each scenario. The minimum number of observations in each node is fixed to 100 and the subsampling fraction is fixed to 0.75. Since the GBM tuning process is computationally intensive, we only tune the GBM parameters for one portfolio per scenario. For the other simulated portfolios within a scenario we rely on these tuned parameters. While the GBM for modelling the layer \texttt{payment} uses a maximum interaction depth larger than two for each scenario, the one to model the layer \texttt{size} uses a maximum interaction depth of one. As such, no interaction effects are used for modelling the average payment sizes.
\begin{table}[ht!]
\centering
\scalebox{0.83}{\begin{tabular}{@{\extracolsep{4pt}}lccccccccc@{}}
\toprule
& \multicolumn{3}{c}{{\bf settlement}} & \multicolumn{3}{c}{{\bf payment}} & \multicolumn{3}{c}{{\bf size}} \\
Scenario & Trees & Shrinkage & Int. depth & Trees & Shrinkage & Int. depth & Trees & Shrinkage & Int. depth \\ \cline{2-4} \cline{5-7} \cline{8-10}     
\rule{0pt}{4ex}1 & 100 & 0.05 & 2 & 125 & 0.05 & 3 & 50 & 0.3 & 1 \\
2 & 50 & 0.2 & 2 & 100 & 0.05 & 5 & 225 & 0.1 & 1 \\
3 & 225 & 0.05 & 1 & 125 & 0.05 & 3 & 700 & 0.05 & 1 \\
4 & 125 & 0.05 & 2 & 50 & 0.2 & 2 & 475 & 0.05 & 1 \\
\bottomrule
\end{tabular}}
\caption{Tuned GBM parameters in each layer and for each scenario: the number of trees (iterations) in the GBM $\in \{25,50,75,\ldots, 1\ 000\}$, the shrinkage parameter $\in \{0.05,0.1,0.2,0.3\}$ and the maximum interaction depth of each tree in the GBM $\in \{1,2,3,\ldots, 7\}$. The subsampling ratio (\texttt{bag.fraction}) is fixed to 0.75 and the minimum number of observations in each node to 100.}
\label{table:tunedcov}
\end{table}

\begin{figure}[ht!]
\centering
\includegraphics[width = \textwidth]{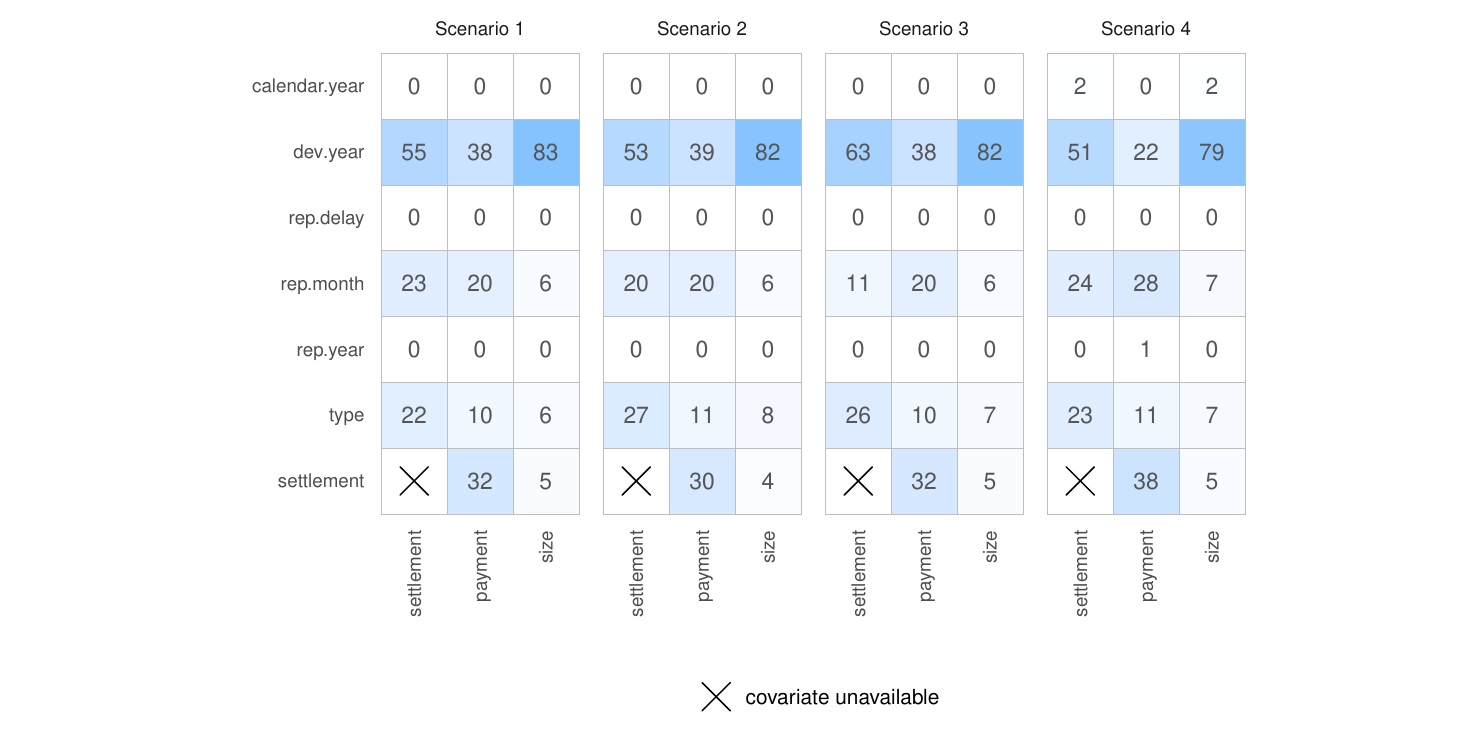}
\caption{The relative variable importance of the covariates in the hierarchical GBM averaged over the 100 simulated data sets of each scenario. The cross-sign indicates that the covariate is not available. \label{fig:importance_gbm} }
\end{figure}

For the four considered scenarios, Figure~\ref{fig:importance_gbm} displays the available covariates together with their relative variable importance in each layer of the hierarchical GBM. We do not include the interaction effect $\texttt{dev.year} * \texttt{rep.month}$ (as revealed by the fitted GLMs, pictured in Figure~\ref{fig:importance_glm}) since GBMs are capable of automatically capturing any relevant interaction effects. Again, not many differences are observed across the scenarios. A claim's reporting month is important to predict whether the claim will settle in the current development year since reporting in the hierarchical GBM, while it was not a selected covariate in the layer \texttt{settlement} of the hierarchical GLM. In the layer \texttt{payment}, the development year since reporting and the reporting month are two additional examples of covariates that matter in the hierarchical GBM, while having limited importance in the calibrated GLMs. These two marginal effects are (somewhat) accommodated by the interaction effect $\texttt{dev.year} * \texttt{rep.month}$ in the hierarchical GLM. Furthermore, the development year since reporting is clearly the main driver in predicting the average payment size in the hierarchical GBM while this is the interaction effect $\texttt{dev.year} * \texttt{rep.month}$ in the hierarchical GLM. However, the GBMs used to model the average payment size have a tuned interaction depth of one, and therefore no interaction effects can be captured within these GBMs. This results in a smaller importance for the predictor \texttt{rep.month} in the GBMs.

\paragraph{Chain ladder.} For each scenario and for each layer, we apply the chain ladder method on the yearly aggregated data per simulated portfolio. Hereto, we aggregate the data per reporting and development year since reporting. For the settlement layer, we focus on the total number of claims that are still open at the beginning of each development year. In the first calendar year following the evaluation date, i.e.~the year 2021, the number of open claims at the start of the calendar year is known, because we register whether or not an individual claim settles in calendar year 2020. Table~\ref{tab:triangle_open} visualizes the resulting run-off triangle for the total number of open claims in a portfolio simulated along the baseline scenario. Note the observed (2020,2) cell with the number of claims reported in the year 2020 that are still open at the start of the year 2021. Our approach to calibrate the chain ladder method for the number of open claims slightly deviates from the traditional setting. As an example, the first development factor in the chain ladder applied to the triangle in Table~\ref{tab:triangle_open} is the ratio between the sum of the values in the red box and the sum of the values in the blue box. We then use this factor to predict the number of claims that are still open at the start of the third development year since reporting, from the set of claims that were reported in the year 2020.

\begin{table}[!ht]
\centering
\begin{tabular}{llccccccccc}
\toprule
& \multicolumn{9}{c}{\textbf{\hspace{2.5cm} Development year since reporting}} \\
\multirow{9}{*}{\rotatebox[]{90}{\textbf{\hspace{-1cm} Reporting year}}} &
 & \textbf{1} & \textbf{2} & \textbf{3} & \textbf{4} & \textbf{5}& \textbf{6} & \textbf{7} & \textbf{8} & \textbf{9} \\
& \textbf{2012} & 11\ 278 & \tikzmark{top left 1}9\ 508 & \tikzmark{top left 2}6\ 634 & 4584 & 3088 & 2055 & 1370 & 908 & 591 \\ 
& \textbf{2013} & 11\ 507 & 9\ 669 & 6\ 812 & 4\ 681 & 3\ 243 & 2\ 175 & 1\ 476 & 949 & 627 \\ 
& \textbf{2014}  & 11\ 248 & 9\ 532 & 6\ 688 & 4\ 551 & 3\ 137 & 2\ 127 & 1\ 421 & 939 & - \\ 
& \textbf{2015}  & 11\ 292 & 9\ 511 & 6\ 635 & 4\ 654 & 3\ 187 & 2\ 173 & 1\ 425 & - & - \\ 
& \textbf{2016}  & 11\ 483 & 9\ 687 & 6\ 710 & 4\ 620 & 3\ 201 & 2\ 142 & - & - & - \\ 
& \textbf{2017}  & 1\ 1291 & 9\ 513 & 6\ 724 & 4\ 671 & 3\ 150 & - & - & - & - \\ 
& \textbf{2018}  & 11\ 226 & 9\ 494 & 6\ 705 & 4\ 675 & - & - & - & - & - \\ 
& \textbf{2019}  & 11\ 410 & 9\ 602 \tikzmark{bottom right 1} & 6\ 686\tikzmark{bottom right 2} & - & - & - & - & - & - \\ 
& \textbf{2020}  & 11\ 597 & 9\ 784 & - & - & - & - & - & - & - \\ 
\bottomrule
\end{tabular}
\caption{The number of open claims aggregated per reporting year and development year in one portfolio simulated along the baseline scenario, with data up to the evaluation date December 31, 2020.}
\label{tab:triangle_open}
\DrawBox[ultra thick, draw=blue, dotted, fill=blue!15, fill opacity=0.3]{top left 1}{bottom right 1}
\DrawBox[ultra thick, draw=red, dotted, fill=red!15, fill opacity=0.3]{top left 2}{bottom right 2}
\end{table}

\subsection{Unraveling the marginal effects on the prediction targets}
We use partial dependence plots to acquire some insights in the marginal effects of the different covariates on the prediction targets in the three different layers (\texttt{settlement}, \texttt{payment}, \texttt{size}). We also check whether the calibrated effects are in line with the effects implemented in the simulation machine. We do this by means of partial dependence plots for those predictors with a variable importance larger than 5 in the calibrated hierarchical GBM in the baseline scenario (see Figure~\ref{fig:importance_gbm}).

\begin{figure}[ht!]
\centering
\includegraphics[width = \textwidth]{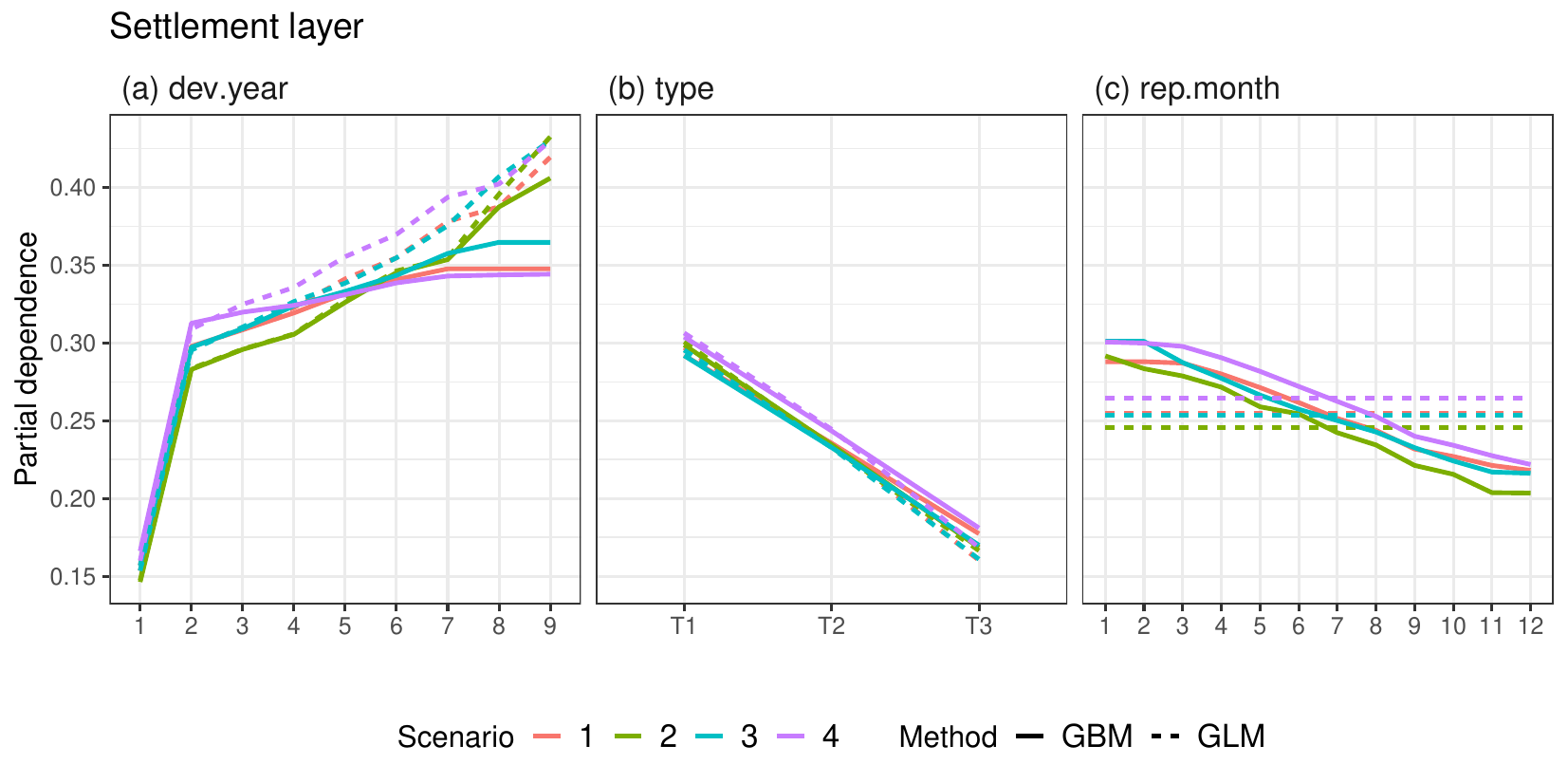}
\caption{Partial dependence plots visualizing the marginal effect of the predictors \texttt{dev.year}, \texttt{type} and \texttt{rep.month} on the settlement indicator in both the hierarchical GLM and hierarchical GBM. The results are based on one portfolio simulated along each of the scenarios specified in Section~\ref{sec:intro.scenarios}.\label{fig:pdpsettlement}}
\end{figure}

\paragraph{Layer settlement.} Figure~\ref{fig:pdpsettlement} displays the partial dependence functions of the predictors \texttt{dev.year}, \texttt{type} and \texttt{rep.month} in the settlement layer of both the calibrated GLM and GBM hierarchical models. As is clear from panel (a), the settlement probability increases in later development years since reporting. In addition, panel (b) indicates that the settlement probability of Type 1 claims is the highest, while it is the lowest for Type 3 claims. From panel (c), the settlement probability of a claim reported towards the end of the year is lower than that of a claim reported at the beginning of the year, at least according to the calibrated hierarchical GBM. We observe a flat partial dependence curve for the hierarchical GLM because the predictor \texttt{rep.month} was not selected during the covariate selection procedure. 

Panels (a) and (b) are in line with the modelling steps in the simulation machine because the settlement delay ratio\footnote{The settlement delay ratio is the actual settlement delay in days divided by the maximal assumed settlement delay in the portfolio.} of a claim is drawn from a beta distribution with \texttt{type}-dependent parameters. First, these parameters are chosen (see Appendix~\ref{appendix:sim.machine}) such that a claim of Type 1 has the lowest expected settlement delay and a claim of Type 3 the highest as captured in panel (b) of Figure~\ref{fig:pdpsettlement}. Second, the parameters in the beta distribution are chosen such that the density of the beta distribution for the settlement delay ratio of a claim is strictly decreasing on the interval $[0,1]$ (in correspondence with the increasing pattern in panel (a) in Figure~\ref{fig:pdpsettlement}). At last, panel (c) is in line with our expectations because the first development year for claims that are reported early in the year is longer than the one for claims reported towards the end of the year. Therefore, as an example, the probability of settlement in the first development year is higher for a claim reported early in the year. This effect is not captured by the hierarchical GLM.

\begin{figure}[ht!]
\centering
\includegraphics[width = \textwidth]{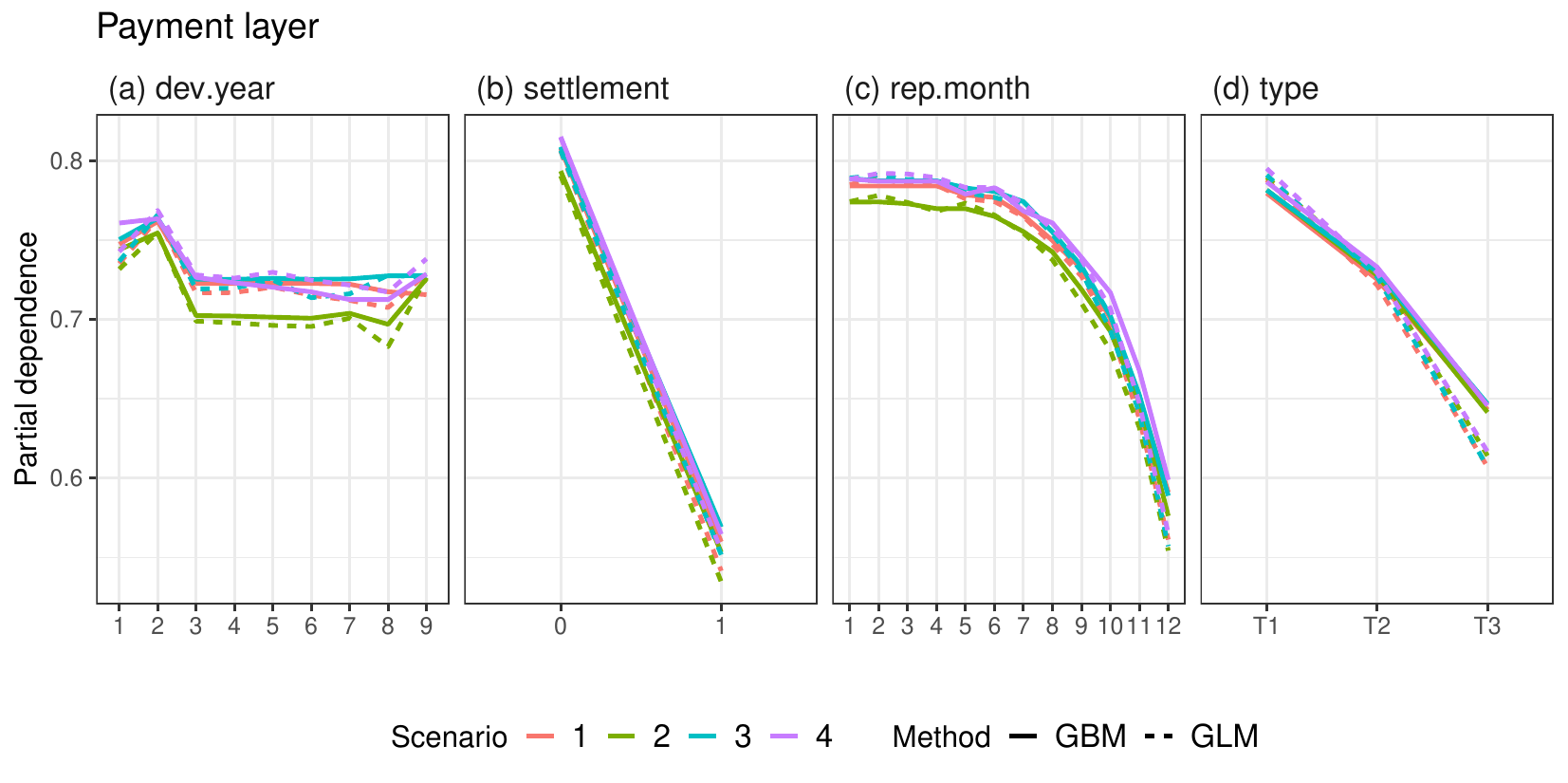}
\caption{Partial dependence plots visualizing the marginal effect of the predictors \texttt{dev.year}, \texttt{settlement}, \texttt{rep.month} and \texttt{type} on the payment indicator in both the hierarchical GLM and hierarchical GBM and this for each scenario.\label{fig:pdppayment}}
\end{figure}

\paragraph{Payment layer.} Figure~\ref{fig:pdppayment} shows the partial dependence functions of the predictors \texttt{dev.year}, \texttt{settlement}, \texttt{rep.month} and \texttt{type} in the layer \texttt{payment} of both the hierarchical GLM and GBM. Panel (a) indicates that the probability of a payment is higher in the second than in the first development year since reporting and decreases again in later development years. In the last development year since reporting, the estimated payment probability again slightly increases. In addition, panel (b) reveals that a payment is less likely in the year of settlement. From panel (c), a payment is less likely to occur for claims reported later in the year. Panel (d) shows that the payment probability is the highest for claims of Type 1 and the lowest for claims of Type 3.

In the simulation machine, we generate the continuous time between two consecutive payments by taking random draws from an exponentially distributed random variable with the rate parameter only depending on the type of the claim. This rate parameter is the highest for claims of Type 1 and the lowest for claims of Type 3, which is in line with the findings in panel (d). In addition, the first payment occurs faster than the subsequent payments (if any) in the continuous time portfolio. We can explain the pattern in panel (a) by the fact that the first development year since reporting is typically not a full calendar year. Together with the fact that the first payment is made more quickly, this leads to an increase in the payment probability in the second development year since reporting and then to a decrease again. Panel (c) aligns with the simulation machine's modelling steps, as the payment probability in the first development year since reporting sharply decreases for claims reported towards the end of the year. At last, panel (b) is in line with our intuitions because the year in which the claim settles is usually shorter than a full calendar year and the payment probability in that year is therefore smaller. 

\begin{figure}[ht!]
\centering
\includegraphics[width = \textwidth]{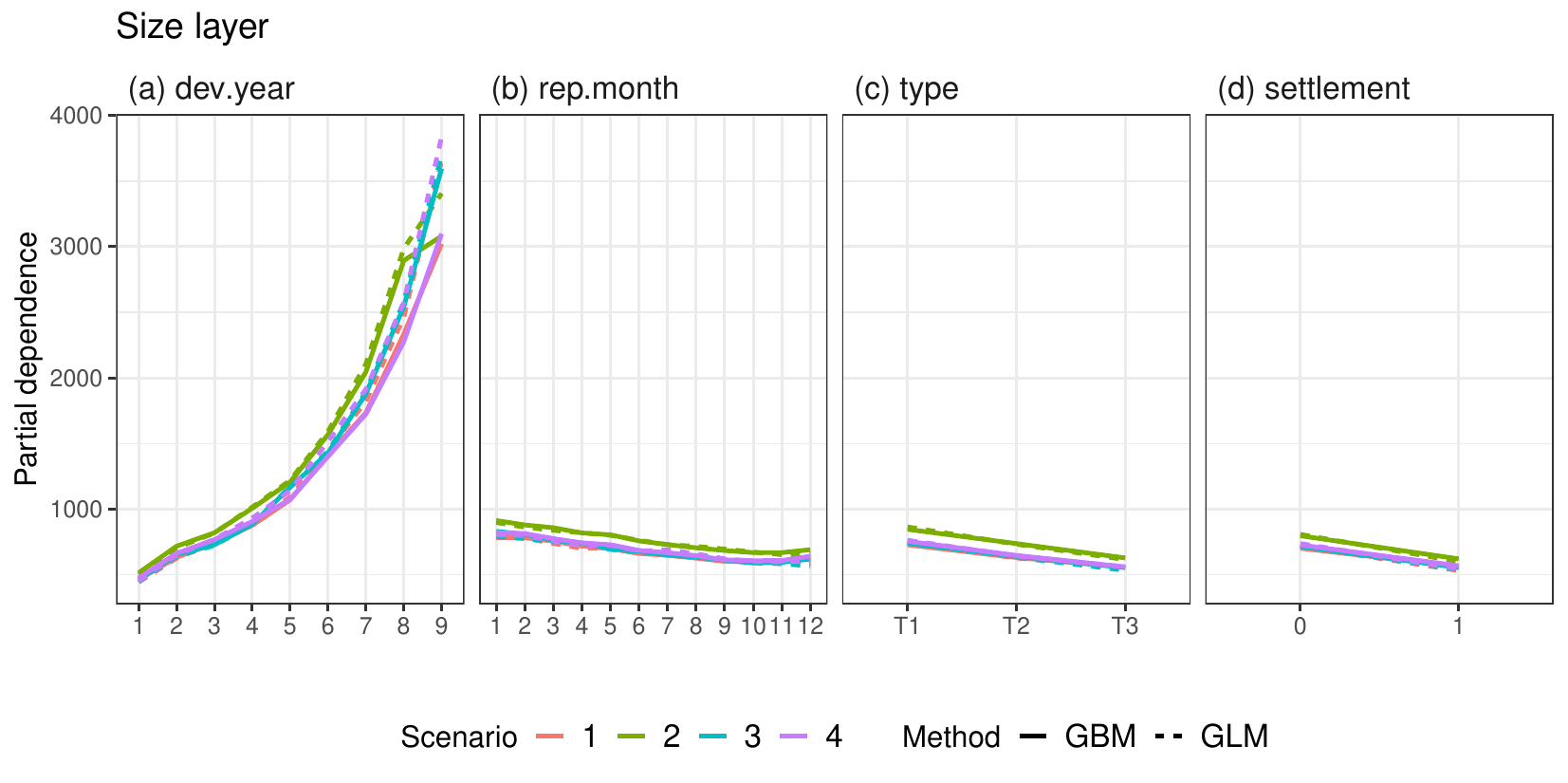}
\caption{Partial dependence plots visualizing the marginal effect of the predictors \texttt{dev.year}, \texttt{rep.month}, \texttt{type} and \texttt{settlement} on the payment size in both the hierarchical GLM and hierarchical GBM and this for each scenario.\label{fig:pdpsize}}
\end{figure}

\paragraph{Size layer.} The marginal effects of the predictors \texttt{dev.year}, \texttt{rep.month}, \texttt{type} and \texttt{settlement} on the payment size are displayed in Figure~\ref{fig:pdpsize} for each scenario and for both the hierarchical GLM and GBM. The predictor \texttt{dev.year} is clearly the main driver behind the predictions for the payment sizes. Panel (a) reveals that payments made in later development years are larger on average. Panel (b) shows that the payment sizes for claims that are reported to the insurer early in the year are also larger on average. The same holds true for claims of Type 1 (panel (c)) and for payment instalments that are not paid in the claim's year of settlement (panel (d)). 

Indeed, the simulation machine generates payment sizes (in the continuous time setting) as random draws from a log-normal distribution where the mean parameter depends on the type of the claim as well as on the continuous time elapsed since reporting.\footnote{To be completely correct, the generated payment sizes also depend on the covariate \texttt{Hidden}. However this is not relevant here as this covariate is not observed by the insurer.} This explains panels (a) and (c). Panel (b) aligns with the modelling set-up as well because the continuous time since reporting is longer for claims reported early in the year. Panel (d) can be explained by the fact that the year of settlement usually does not last a full calendar year. As a result, fewer payments in the continuous time portfolio can be made and hence, the payment size in the year of settlement is lower.

\subsection{Restoring the balance property}
We apply the bias correction step as explained in Section~\ref{section:covariateselection}. Hereto we calculate the development year specific bias correction factors for each portfolio and each layer, both for the hierarchical GLM and GBM. Table~\ref{tab:sim:biasregfac} shows the results averaged over the 100 portfolios simulated along the baseline scenario. In the layer \texttt{payment} of the hierarchical GLM, the bias regularization factors are equal to one because of the balance property in a logistic regression model. In the layer \texttt{settlement} and to a lesser extent in the layer \texttt{size}, the bias correction factors are also very close to one. We observe some larger deviations from one in the hierarchical GBM at later development years since reporting. In the layer \texttt{payment}, we usually overestimate the actual amount of payments in the training set, leading to a bias correction factor that is less than one in most development years. The opposite is true for the layers \texttt{settlement} and \texttt{size} in the hierarchical GBM.
\begin{table}[!ht]
\centering
\begin{tabular}{llccccccccc}
\toprule
& \multicolumn{9}{c}{\textbf{\hspace{2.5cm} Development year since reporting}} \\
& & \textbf{1} & \textbf{2} & \textbf{3} & \textbf{4} & \textbf{5}& \textbf{6} & \textbf{7} & \textbf{8} & \textbf{9} \\
\multirow{3}{*}{\rotatebox[]{90}{\textbf{GLM}}}  & settlement & 1.000 & 1.000 & 1.000 & 1.000 & 1.000 & 1.000 & 1.000 & 1.000 & 0.999 \\ 
 & payment & 1.000 & 1.000 & 1.000 & 1.000 & 1.000 & 1.000 & 1.000 & 1.000 & 1.000 \\ 
 & size & 0.996 & 1.000 & 1.000 & 0.998 & 0.999 & 1.000 & 1.002 & 1.002 & 1.006 \\ \\
\multirow{3}{*}{\rotatebox[]{90}{\textbf{GBM}}} & settlement & 0.959 & 0.999 & 1.007 & 1.015 & 1.023 & 1.039 & 1.077 & 1.131 & 1.196 \\ 
 & payment & 1.003 & 1.008 & 0.993 & 0.990 & 0.988 & 0.985 & 0.983 & 0.980 & 0.979 \\ 
 & size & 0.997 & 1.000 & 1.005 & 1.013 & 1.014 & 1.027 & 1.043 & 1.071 & 1.215 \\ 
\bottomrule
\end{tabular}
\caption{The bias correction factors per development year since reporting in both the hierarchical GLM and the hierarchical GBM and averaged over the 100 portfolios simulated along the baseline scenario.\label{tab:sim:biasregfac}}
\end{table}

\subsection{Prediction results and discussion}\label{sec:predictionresults}
For each of the four considered scenarios, we predict the expected number of open claims, the expected number of payments and the expected payment sizes (RBNS reserve) in each calendar year following the evaluation date December 31, 2020. We do this for those claims reported reported between January 1, 2012 and December 31, 2020 and track the claims up to 9 development years after reporting (see Figure~\ref{tikz:structuredata}). We then add up these predictions for each layer.\footnote{We know the exact number of open claims in calendar year 2021 and therefore omit this year for evaluating the predictions of the expected number of open claims after the evaluation date.} In addition, we compare these aggregated predictions with the ones obtained with the chain ladder method, as explained in Section~\ref{section:calibrationhrm}. Further, we measure the predictive performance of the reserving models through the percentage error, defined in~\eqref{eq:PE.error}, where the observed number of open claims, payments and payment sizes after the evaluation date can be retrieved from each of the simulated portfolios.

\begin{figure}[ht!]
\centering
\includegraphics[width = 0.84\textheight, height = \textwidth, angle = 90]{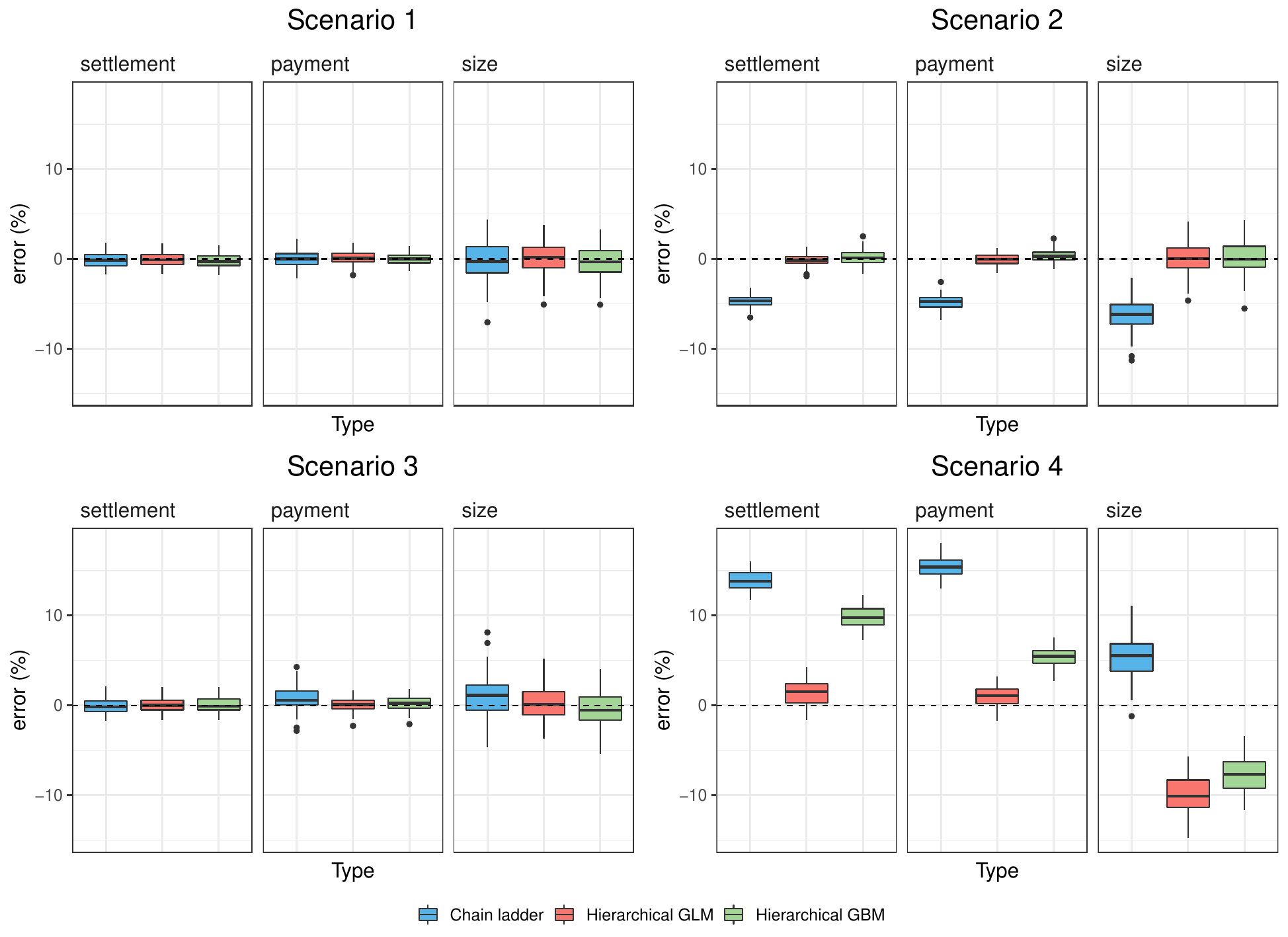}
\caption{The boxplots showing the percentage errors of the predicted number of open claims, the predicted number of claim payments and the predicted payment sizes aggregated over the eight calendar years following the evaluation date December 31, 2020. The percentage errors are based on the results from 100 portfolios simulated along each scenario. We show the results for the chain ladder method (blue), the hierarchical GLM (red) and the hierarchical GBM (green). \label{fig:boxplotsall}}
\end{figure}

\paragraph{Scenario 1: Baseline scenario.} The upper left panel in Figure~\ref{fig:boxplotsall} shows the boxplots of the percentage errors in each layer for the three reserving models, namely the chain ladder method, the hierarchical GLM and the hierarchical GBM, calibrated on the 100 portfolios simulated along the baseline scenario. The three reserving models are capable of predicting the relevant outcome variables reasonably well. All three models have a similar percentage error distribution.

\paragraph{Scenario 2: Claim mix scenario.} The upper right panel in Figure~\ref{fig:boxplotsall} shows the predictive model performances in each layer of the considered reserving models, calibrated on each of the 100 portfolios simulated along the claim mix scenario. The hierarchical reserving models perform very well as if there was no change in claim mix. The chain ladder method underestimates the actual number of open claims, the number of payments and the payment sizes. Since the settlement probability, payment probability and payment sizes all depend on the claim type, changes in claim mix violate the stability assumptions underlying the chain ladder method. Although part of the impact of this disturbance is captured by the reporting year specific parameters, the impact on the overall reserve remains large. However, the performance of the chain ladder method can be improved by building claim type specific chain ladder methods. 

\paragraph{Scenario 3: Extreme event scenario.} Figure~\ref{fig:boxplotsall} (bottom left panel) shows the performance of the predictive models when calibrated on the extreme event portfolios. All three reserving models perform quite well in each layer. A slightly worse performance is observed for the chain ladder method when predicting the number of payments and to a lesser extent when predicting the payment sizes.

\begin{figure}[ht!]
\centering
\includegraphics[width = 0.9\textwidth]{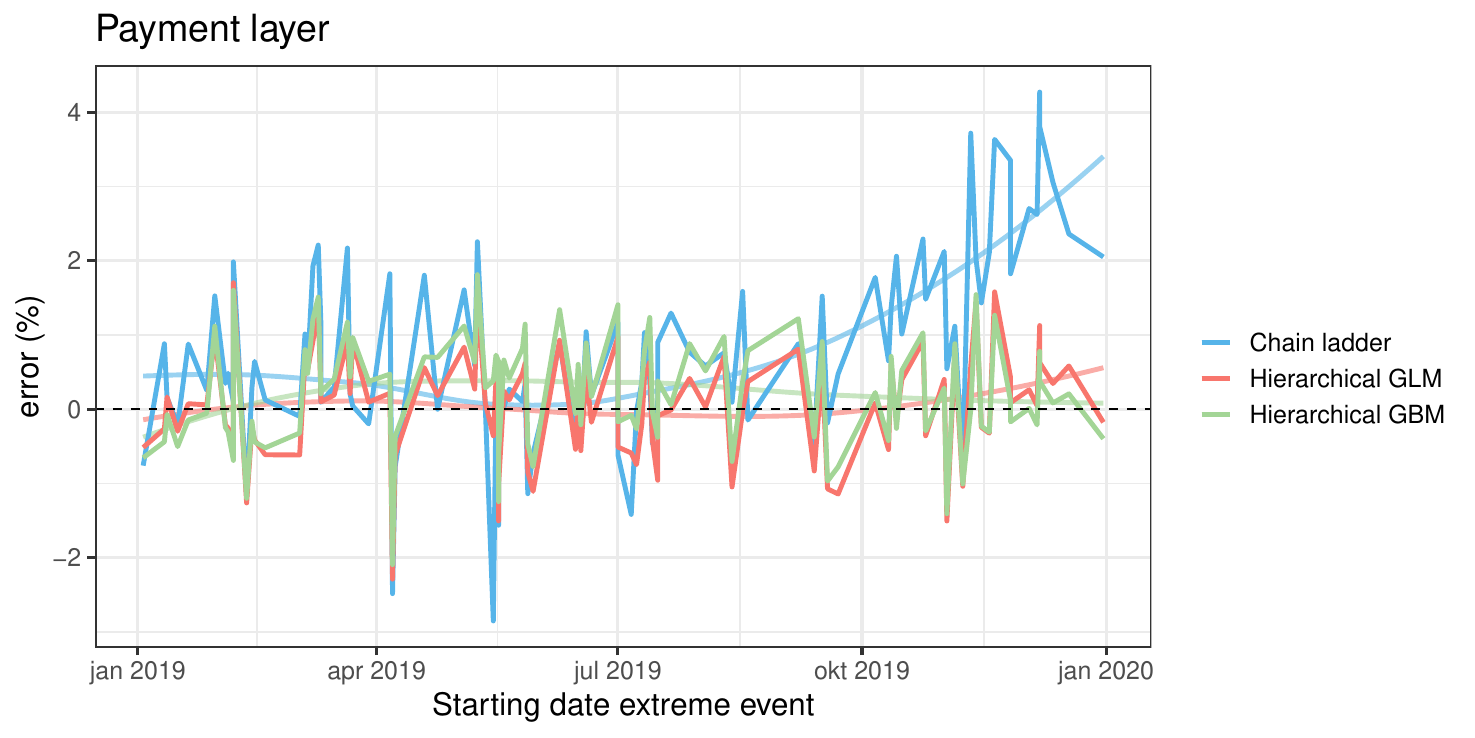}
\caption{The percentage error of the predicted number of payments that occur after the evaluation date December 31, 2020 in the 100 portfolios simulated along the extreme event scenario, as a function of the start date of the extreme event. The shaded line shows a smooth approximation through the percentage errors.\label{fig:layer_payment_startdate}}
\end{figure}

Figure~\ref{fig:layer_payment_startdate} displays the percentage error  of the predicted number of payments as a function of the starting date (within the year 2019) of the extreme event. We show the results for each of the 100 portfolios simulated along this third scenario. The hierarchical GLM and the hierarchical GBM perform equally well regardless of the starting date of the extreme event. The chain ladder method performs worse when the extreme event occurs near the end of the year. We can intuitively explain this by taking a closer look at the development process of claims that occurred during an extreme event towards the end of the year 2019. Many of these claims will then be reported in early 2020. Since the first reporting year runs from the reporting date until December 31 of the same year, claims reported around the start of the calendar year typically have a higher payment probability than those reported later in the year. This is also confirmed by panel (c) of Figure~\ref{fig:pdppayment}. As a result, their number of payments in the first development year will be significantly larger than what is expected based on the historical payment pattern in the first development year. However, this is not representative for the number of payments in later years. Consequently, when the chain ladder method applies the same development factor, calibrated on the 2012-2019 data, to the 2020 row, this results in an overestimation of the reserve. Due to this overestimation of the number of payments, the prediction of the total payment sizes is also too high. Note that the individual hierarchical reserving models perform better because \texttt{rep.year} was not selected. Hence these models correctly predict that for later development years of 2020 the number of payments per open claim does not significantly differ from previous years.

\paragraph{Scenario 4: Change in settlement delay.} The bottom right panel in Figure~\ref{fig:boxplotsall} shows the boxplots of the percentage errors in the fourth scenario. The chain ladder method severely overestimates the number of open claims, the number of payments and the total payment sizes in the eight calendar years following the year 2020 in the portfolios. The same holds true for the hierarchical GBM but to a slightly lesser extent. The hierarchical GLM leads to a reasonably well predictive performance except for the layer \texttt{size}. We get these poor predictions for the total payment sizes in the simulated portfolios because the development year since reporting has a different effect on the average payment size for claims that occur before and after the year 2017. Hence, there is a significant interaction term \texttt{rep.year:dev.year} present that affects the average payment sizes, which cannot be captured by the constructed reserving models because of the limited available training data to learn this effect.

\section{Conclusion}
We propose the hierarchical reserving model as a general framework for RBNS reserving in discrete time. By adding layers and choosing predictive models this framework can be tailored to a wide range of claim development scenarios. At the same time, our approach enables the development of best practices for calibration. In addition, model comparison becomes easier since the ideas implemented in many existing reserving models can be rephrased to hierarchical reserving models. Moreover, we establish a connection between the hierarchical reserving model that uses data registered at the level of individual claims and some aggregate reserving models. This allows for a data driven choice between aggregate and individual reserving. We illustrate our framework on a detailed case study with a home insurance data set. The flexibility of the framework is demonstrated by calibrating the same three layer structure with generalized linear models and gradient boosting models. As a best practice, we minimize the effect of day-to-day volatility to compare our reserving models, by evaluating the predictive performance over 365 evaluation days. The individual hierarchical models outperform the chain ladder method designed for aggregated data and have the additional benefit that extreme events do not have to be removed prior to reserving. The excellent performance of the hierarchical reserving model is confirmed in a simulation study where we evaluate its predictive performance on portfolios simulated along different scenarios such as a change in the claim mix, an extreme event scenario and a change in the settlement delay of the claims.

Future research will put focus on quantifying the prediction uncertainty in the hierarchical reserving models, inspired by the ideas of \citet{gneiting2007strictly}. Next to this, further work regarding the automatic selection of features and the design and inclusion of additional layers (e.g.~to handle negative payments) in the hierarchical reserving model is valuable. 

\section{Acknowledgements}
The authors thank the anonymous referees and the editor for useful comments which led to significant improvements of the paper. This work was supported by the Argenta Research chair at KU Leuven; KU Leuven's research council [project COMPACT C24/15/001]; and Research Foundation Flanders (FWO) [grant number 11G4619N]. Katrien Antonio acknowledges the support of the research chair DIALog, sponsored by CNP Assurances. We thank Roel Verbelen for his helpful comments at the start of this project.  

{
\bibliographystyle{plainnat}
\bibliography{references}}

\appendix 

\section{The simulation machine}\label{appendix:sim.machine}
Inspired by \cite{gabrielli2018individual} and \cite{avanzi2021synthetic} we develop a simulation machine to generate portfolios with the development of individual claims over time. We choose to build our own simulation machine to accommodate tailored adjustments to the different simulation steps. This allows us to easily create scenarios that we can use to investigate the predictive power of our proposed 3-layer hierarchical reserving model. We describe the technicalities behind our simulation machine in this appendix. Our simulation procedure consists of 6 simulation steps and a data wrangling step to go from a data set in a continuous date format to a data set in an annual, discrete time. We provide the details of constructing one portfolio data set along the baseline scenario. 

\subsection{Working principles}
Figure~\ref{tikz:development_individual} illustrates the simulation steps for one claim. In the simulation process, we mimick reality as closely as possible by chronologically modelling the different steps in the claims handling process. Technical details are deferred to Appendix~\ref{subappendix:technicaldetails}. First, $n = 125\ 000$ occurrence dates are simulated between January 1, 2010 and December 31, 2020. Next, the reporting delays are generated with an assumed maximal reporting delay of 2 years. We remove all claims reported before January 1, 2012. Eventually, we obtain a 9-year observation window of reported claims as shown in Figure~\ref{tikz:structuredata}. Per claim we generate 30 payments and the corresponding payment sizes. Next, we simulate the settlement delay. Payments that occur after the settlement date or 9 years after reporting are considered invalid and removed from the portfolio (see Figure~\ref{tikz:development_individual}). At last, we convert the portfolio to an annualized basis in line with the discrete time hierarchical reserving model proposed in this paper. Figure~\ref{tikz:aggregration} visualizes this discretization process. We hereby aggregate the payment sizes of the payments that take place in the same development year since reporting.

\begin{figure}[ht!]
\centering
\begin{scaletikzpicturetowidth}{\textwidth}
\begin{tikzpicture}[snake=zigzag, line before snake = 2mm, line after snake = 2mm, scale=\tikzscale, text width = 1.75cm, align = center]
    \draw[snake] (0,0) -- (1,0);
    \draw        (1,0) -- (3.25,0);
    \draw[snake] (3.25,0) -- (4.25,0);
    \draw        (4.25,0) -- (5.4,0);
    \draw[snake] (5.4,0) -- (6.4,0);
    \draw        (6.4,0) -- (6.6,0);
    \draw[->]    (6.6,0) -- (7.6,0);

    \foreach \x in {1,2,4.8,5.88}
      \draw (\x cm,5pt) -- (\x cm,-5pt);
      
    \foreach \x in {2.2,2.9,3.2,4.25,5.2,6.8}
      \draw (\x cm,2pt) -- (\x cm,-2pt); 

    \draw (1,0)    node[below=15pt] {\small Occurrence date};
    \draw (2,0)  node[below=15pt] {\small Reporting date};
    \draw (4.8,0)  node[below=15pt] {\small Settlement date};
    \draw (5.88,0) node[below=15pt] {\small End \\ date};

	\draw (2.2,0)  node[below=5pt] {\color{ForestGreen}\scriptsize $1$};
	\draw (2.9,0)  node[below=5pt] {\color{ForestGreen}\scriptsize $2$};
	\draw (3.2,0)  node[below=5pt] {\color{ForestGreen}\scriptsize $3$};
	\draw (4.25,0) node[below=5pt] {\color{ForestGreen}\scriptsize $k$};
	\draw (5.2,0)  node[below=5pt] {\color{red} \sout{\scriptsize $k+1$}};
	\draw (6.8,0)  node[below=5pt] {\color{red}\sout{\scriptsize $30$}};

	\draw[<->]   (1.05,0.25) -- (1.95,0.25);
	\draw (1.50,0.25) node[text width = 3cm, above = 2pt] {\scriptsize Reporting delay};
	
	\draw[<->]   (2.05,0.25) -- (4.75,0.25);
	\draw (3.4,0.25) node[text width = 3cm, above = 2pt] {\scriptsize Settlement delay};
	
	\draw[<->] (2.95,-0.22) -- (3.15,-0.22);
	\draw (3.05,-0.22) node[text width = 3cm, below = 2pt] {\scriptsize Payment delay};
	
    \draw (0,0) node[circle,fill,inner sep=0.01pt, text width = 0.1cm]{};
    \draw (0,0) node[above=1pt, text width = 1cm] {\scriptsize Jan.~1};
	\draw (0,0) node[below=1pt, text width = 1cm] {\scriptsize 2010};
	
	\draw (1.45,0) node[circle,fill,inner sep=0.01pt, text width = 0.1cm]{};
	\draw (1.45,0) node[above=1pt, text width = 1cm] {\scriptsize Jan.~1};
	\draw (1.45,0) node[below=1pt, text width = 1cm] {\scriptsize 2012};
	
	\draw [decorate,decoration={brace,amplitude=5pt,mirror,raise=10ex}, text width = 6cm]
  (1.9,0) -- (5.88,0) node[midway,yshift=-5.5em]{\scriptsize Maximal development of 9 years};
\end{tikzpicture}
\end{scaletikzpicturetowidth}
  \caption{Development of an individual claim from the simulation machine. In this example the claim occurred before January 1, 2012 but is reported afterwards. The numbers in green represent the valid payment dates. The crossed out numbers in red indicate the invalid payment dates occurring after the settlement date. The end date refers to the date nine years after the reporting date since we track claims over a maximum of nine years. \label{tikz:development_individual}}
\end{figure}

\begin{figure}[ht!]
\centering
\begin{scaletikzpicturetowidth}{\textwidth}
\begin{tikzpicture}[snake=zigzag, line before snake = 2mm, line after snake = 2mm, scale=\tikzscale, text width = 1.75cm, align = center]
    \draw[->]    (0,0) -- (13,0);

    \foreach \x in {1.75, 3.45, 9.5}
      \draw (\x cm,7.5pt) -- (\x cm,-7.5pt);
      
    \foreach \x in {0,1,2,...,12}
          \draw (\x,0) node[circle,fill,inner sep=0.01pt, text width = 0.1cm]{};
          
          
     \foreach \x in {2010,2011,2013,2014,...,2022}    
          \draw (\x-2010,0) node[below=1pt, text width = 1cm] {\scriptsize \x}; 
          
     \draw (2.1,0) node[below=1pt, text width = 1cm] {\scriptsize 2012};     
          
     \foreach \x in {2,3,...,6}      
          \draw [decorate,decoration={brace,amplitude=5pt,mirror,raise=3ex}, text width = 1cm]
  (\x+2.05,0) -- (\x+2.95,0) node[midway,yshift=-2.5em]{\scriptsize \x}; 
  
     \draw [decorate,decoration={brace,amplitude=5pt,mirror,raise=3ex}, text width = 1cm]
  (3.45,0) -- (4,0) node[midway,yshift=-2.5em]{\scriptsize 1}; 

     \draw [decorate,decoration={brace,amplitude=5pt,mirror,raise=3ex}, text width = 1cm]
  (9,0) -- (9.5,0) node[midway,yshift=-2.5em]{\scriptsize 7}; 
            
    \draw (1.75,0) node[above=10pt] {\small Occurrence date};
    \draw (3.45,0) node[above=10pt] {\small Reporting date};
    \draw (9.5,0) node[above=10pt] {\small Settlement date};
    \draw (12,0)   node[above=10pt] {\small End \\ date};

\end{tikzpicture}
\end{scaletikzpicturetowidth}
  \caption{Transformation of a claim from a continuous to a yearly, discrete time framework. The dots indicate the start date (January 1) of each year. The claim occurs in occurrence year 2011 and is reported in year 2013 to the insurer. The horizontal curly brackets denote the development years since reporting. The claim settles in its seventh development year since reporting. We aggregate the payment sizes that take place in the same development year since reporting. \label{tikz:aggregration}}
\end{figure}

\subsection{Technical details}\label{subappendix:technicaldetails}

\paragraph{Step 1: Simulating the occurrence date.} In the baseline scenario, we assume that occurrence dates are uniformly distributed on a time interval $\left[t_{\min}^o,t_{\max}^o\right]$, where $t_{\min}^o$ is the minimal occurrence date, i.e.~January 1, 2010, and $t_{\max}^o$ the maximum occurrence date, i.e.~December 31, 2020. The occurrence date random variable $A$ thus has the following distribution:
$$ \mathbb{P}(A = t) = \dfrac{1}{\text{days}(t_{\min}^o, t_{\max}^o)}, \hspace{1cm} t \in \left[t_{\min}^o, t_{\max}^o\right], $$
where $t$ represents a continuous date between $t_{\min}^o$ and $t_{\max}^o$ and where $\text{days}\left(t_{\min}^o, t_{\max}^o\right)$ is the number of days in the interval $\left[t_{\min}^o, t_{\max}^o\right]$. We simulate $n$ occurrence dates from the imposed uniform distribution. We denote such a realization by $a_k$, representing the occurrence date of a claim $k$.

\paragraph{Step 2: Simulating the covariates.} In this second step, we simulate two 3-level categorical covariates that can be considered as policy characteristics. The first covariate \texttt{type} represents the type of the claim and the second covariate \texttt{hidden} is an unobservable covariate, used to generate the development of a claim but assumed unavailable to the reserving actuary. The assumed discrete sampling distributions for both covariates are:
\begin{align*}
&\mathbb{P}(\texttt{type} = \tau) = p_\tau, \hspace{1.4cm} \tau \in \{\texttt{T1}, \texttt{T2}, \texttt{T3}\}, \\
&\mathbb{P}(\texttt{hidden} = h) = q_h, \hspace{1cm} h \in \{\texttt{L}, \texttt{M}, \texttt{H}\},
\end{align*}
where $p_{\texttt{T1}}, p_{\texttt{T2}}, p_{\texttt{T3}} \geq 0$ as well as $q_L, q_M, q_H \geq 0$ and where both add up to one. In our simulation machine, we impose the discrete probabilities $(p_{T1}, p_{T2}, p_{T3}) = (0.60,0.25,0.15)$ and $(q_L, q_M, q_H) = (0.35,0.45,0.20)$, but they can be chosen freely. Independent of the first simulation step, we draw a sample of size $n$ from both distributions.

\paragraph{Step 3: Simulating the reporting delay.} The maximal assumed reporting delay is two years. Given the type of a claim, we make the following distributional assumption for the random variable (r.v.) $\tilde D^r$ that represents the ratio between the reporting delay $D^r$ of the claim and the maximal reporting delay of two years:
$$ \tilde{D}^r \mid \texttt{type} = \tau \sim \text{Beta}(\alpha_\tau, \beta),$$
where $\tau \in \{\texttt{T1}, \texttt{T2}, \texttt{T3}\}$. We fix $\beta$ to 10 and make the following choices for $\alpha$:
$$ \alpha_{\texttt{T1}} = 1, \hspace{0.2cm} \alpha_{\texttt{T2}} = 2, \hspace{0.2cm} \alpha_{\texttt{T3}} = 3.$$
The density of the r.v. $\tilde{D}^r$ thus equals:
$$ f_3\left(x;\alpha_\tau,\beta\right) = \frac{\Gamma(\alpha_\tau +
\beta)}{\Gamma(\alpha_\tau)\Gamma(\beta)}x^{\alpha_\tau - 1}\left(1 - x\right)^{\beta - 1}, \hspace{0.5cm} x \in (0,1). $$
Conditional on the feature space after the second simulation step, we then simulate reporting delay ratios $\tilde{d}^r_k$ of each claim $k$. The simulation of the actual reporting delay $d^r_k$ is simply obtained by multiplying $\tilde{d}^r_k$ with the maximal reporting delay of two years. The reporting date $r_k$ then equals the sum of the occurrence date $a_k$ and the reporting delay $d^r_k$. We only keep the claims with a reporting date after January 1, 2012.

\paragraph{Step 4: Simulating the payment delays.} For every claim, we simulate 30 payment delays, i.e.~the time between two consecutive claim payments. In a later step, we remove the payments that occur after the claim settlement date or that fall outside the 9-year observation window. The inter-arrival times $T_v$, for $1 \leq v \leq 30$, of these 30 claim payments follow an exponential distribution with a rate parameter $\lambda_{v,\tau}$ depending on the type $\tau$ of the claim. In addition, the first payment occurs at a faster rate compared to the next ones:
\begin{align*}
T_1 \mid \texttt{type} &= \tau \sim \text{Exp}(\lambda_{1,\tau}), \\
T_v \mid \texttt{type} &= \tau \sim \text{Exp}(\lambda_{2,\tau}) \hspace{0.5cm} 2\leq v \leq 30,
\end{align*}
where $\tau \in \{\texttt{T1}, \texttt{T2}, \texttt{T3}\}$ and with the following choices for the rate parameters:
\begin{align*}
(\lambda_{1,\texttt{T1}}, \lambda_{1,\texttt{T2}}, \lambda_{1,\texttt{T3}})  &= (6,5,4) \\
(\lambda_{2,\texttt{T1}}, \lambda_{2,\texttt{T2}}, \lambda_{2,\texttt{T3}})  &= (2,1.5,1).
\end{align*}
We denote the simulated path of payment delays for claim $k$ by $\boldsymbol{d}^p_k = \left(d_{k,1}^p, d_{k,2}^p, \ldots, d_{k,30}^p\right) \in \mathbb{R}^{30}$. The actual payment date vector of claim $k$, denoted as $\boldsymbol{p}_k \in \mathbb{R}^{30}$, then consists of the dates:
\begin{align*}
p_{k,l} = r_k + \displaystyle \sum_{m=1}^l d_{k,m}^p, \hspace{0.5cm} 1\leq l \leq 30.
\end{align*}

\paragraph{Step 5: Simulating the payment sizes.} For every payment date, we simulate a size $y$ from the log-normal distributed payment size r.v.~$Y$. We assume that the mean of this log-normal distribution depends on the time span between the reporting date $r$ and the claim payment date $p$, say $\Delta t_{r \rightarrow p}$, on the type of the claim and on the level of the hidden covariate:
$$ Y \mid \texttt{type} = \tau, \: \texttt{hidden} = h,\: \Delta t_{r \rightarrow p} \sim \text{LN}\left(\mu\left(\tau,h,\Delta t_{r \rightarrow p}\right), \sigma^2\right),$$
where the variance $\sigma^2$ is fixed to 1 and where the mean equals
\begin{align*}
\mu\left(\tau, h,\Delta t_{r \rightarrow p}\right) &= \log \gamma_\tau + 0.1 \cdot \left(\Delta t_{r \rightarrow p}\right)^{\delta_h},
\end{align*}
with the following choices for the covariate-dependent parameters:
\begin{align*}
(\gamma_{\texttt{T1}}, \gamma_{\texttt{T2}},\gamma_{\texttt{T13}}) &= (100,200,400), \\
(\delta_{\texttt{L}}, \delta_{\texttt{M}}, \delta_{\texttt{H}}) &= (1.50,1.25,1.40).
\end{align*}
The log-normal density then equals
$$f_6\left(y;\mu\left(\tau,h,\Delta t_{r \rightarrow p}\right),\sigma^2\right)  = \dfrac{1}{y\sigma \sqrt{2\pi}} \exp\left\{-\dfrac{1}{2}\left(\dfrac{\log (y / \gamma_\tau) - 0.1 \cdot \left(\Delta t_{r \rightarrow p}\right)^{\delta_h}}{\sigma}\right)^2\right\}, \hspace{0.5cm} x > 0.$$
For every claim $k$ and for each simulated claim payment date $p_{k,l}$ in Step 5, we then simulate a claim payment size $y_{k,l}$ from the imposed log-normal distribution. The result is a 30-dimensional vector of claim payment sizes $\boldsymbol{y}_{k} \in \mathbb{R}^{30}$. 

\paragraph{Step 6: Simulating the settlement delay.} The maximal assumed settlement delay is 20 years. The simulation strategy used for settlement delays is similar to that of the reporting delay. Given the type of a claim, we assume that the ratio $\tilde D^s$ of the settlement delay r.v. $D^s$ and the maximal settlement delay of 20 years follows a $\text{Beta(}\alpha, \beta_j\text{)}$ distribution :
$$ \tilde{D}^s \mid \texttt{type} = \tau \sim \text{Beta}(\alpha, \beta_\tau),$$
where we now fix $\alpha$ to 1 and where we make the following choices for $\beta$:
$$ \beta_{\texttt{T1}} = 8, \hspace{0.2cm} \beta_{\texttt{T2}} = 6, \hspace{0.2cm} \beta_{\texttt{T3}} = 4.$$
Conditional on the feature space after the third simulation step, we simulate settlement delay ratios $\tilde{D}^s_k$ of each claim $k$. The simulation of the actual settlement delay $d^s_k$ is obtained by multiplying $\tilde{d}^s_k$ with the maximal settlement delay of 20 years. The settlement date $s_k$ then equals the sum of the reporting date $r_k$ and the settlement delay $d^s_k$. 

We also set the simulated claim payment size $y_{k,l}$ to \texttt{NA} whenever the claim payment date $p_{k,l}$ falls outside the 9-year observation window or occurs after the claim settlement date $s_k$.

\paragraph{Step 7: Data wrangling.} In this last step, we first move from a continuous time setting to a discrete time setting on an annual basis. We stick to an individual claim focus and only perform an aggregation in the time dimension. Let $year(\cdot$) and $month(\cdot)$ be the functions that return the year and the month respectively, from an input date `yyyy/mm/dd'. We then define the occurrence year, the occurrence month, reporting year and reporting month as:
\begin{align*}
\texttt{occ.year}_k &= year(a_k), & \texttt{occ.month}_k &= month(a_k),\\
\texttt{rep.year}_k &= year(r_k), &\texttt{rep.month}_k &= month(r_k).
\end{align*}
Next, we apply the $year(\cdot)$ function to each component of the payment date vector $\boldsymbol{p}_{k}$ and aggregate those payment sizes $\boldsymbol{y}_k$ of a claim $k$ occurring in the same year. So for each claim $k$, we obtain a 9-dimensional payment size vector $\boldsymbol{y}_k^{\text{Dev}} \in \mathbb{R}^9$ that represents the total payment size of claim $k$ in every development year $j \in \{1,2,\ldots, 9\}$ since reporting. In addition we define a payment indicator vector $\text{\textbf{pay}}_k^{\text{Dev}}$ that indicates whether a payment for claim $k$ is made in development year $j$ since reporting.

We define two new claim characteristics for every claim $k$, namely a settlement and open indicator. The settlement indicator is a 9-dimensional vector that indicates whether a claim $k$ is settled at the end of each development year $j \in \{1,2,\ldots, 9\}$ since reporting. First, let $\nu_k$ be the number of years that the claim $k$ is open after the reporting date $\nu_k$:
$$ \nu_k = \min (9, \text{year}(s_k) - \text{year}(r_k) + 1).$$
Then we define the settlement indicator \textbf{cl} as:
$$ \textbf{\text{cl}}^{\text{Dev}} = (\underbrace{0,\ldots,0}_{\nu_k - 1},\underbrace{1,\ldots,1}_{9 - \nu_k + 1}) \in \mathbb{R}^9.$$
The open indicator \textbf{op} indicates whether the claim $k$ is still open at the beginning of each development year $j \in \{1,2,\ldots, 9\}$ since reporting and is therefore equal to:
$$ \textbf{\text{op}}^{\text{Dev}} = (\underbrace{1,\ldots,1}_{\nu_k},\underbrace{0,\ldots,0}_{9 - \nu_k}) \in \mathbb{R}^9.$$

In a last step, we expanded our data set based on the development year since reporting. This results in 9 observations per claim $k$.

\subsection{Example of a simulated claim.} 
Table~\ref{tab:overviewsimulationmachine} shows an overview of the characteristics of one claim, simulated along the working principles of the baseline scenario. The claim occurs at October 13, 2012 (occurrence year 2012). It is a Type 2 claim and the level of the unobservable covariate \texttt{hidden} equals \texttt{M}. The claim is reported to the insurer $0.1290$ years (80 days) after occurrence (reporting year 2013). The claim settles at October 24, 2015 (calendar year 4). Hence, the claim settles in the third development year since reporting. In the first three development years, the aggregated payment sizes are $1030.93$, $1353.22$ and $518.00$.

\begin{table}[h!]
\centering
\scalebox{0.65}{\begin{tabular}[t]{llllllllll} 
\toprule
Development year & 1 & 2 & 3 & 4 & 5 & 6 & 7 & 8 & 9 \\
\hline 
\toprule
claim.nr & 1 & 1 & 1 & 1 & 1 & 1 & 1 & 1 & 1 \\ 
  type & T2 & T2 & T2 & T2 & T2 & T2 & T2 & T2 & T2 \\ 
  hidden & M & M & M & M & M & M & M & M & M \\ 
  occ.date & 2012/10/13 & 2012/10/13 & 2012/10/13 & 2012/10/13 & 2012/10/13 & 2012/10/13 & 2012/10/13 & 2012/10/13 & 2012/10/13 \\ 
  occ.year & 2012 & 2012 & 2012 & 2012 & 2012 & 2012 & 2012 & 2012 & 2012 \\ 
  occ.month & 10 & 10 & 10 & 10 & 10 & 10 & 10 & 10 & 10 \\ 
  rep.delay & 0.2190 & 0.2190 & 0.2190 & 0.2190 & 0.2190 & 0.2190 & 0.2190 & 0.2190 & 0.2190 \\ 
  rep.date & 2013/01/01 & 2013/01/01 & 2013/01/01 & 2013/01/01 & 2013/01/01 & 2013/01/01 & 2013/01/01 & 2013/01/01 & 2013/01/01 \\ 
  rep.year & 2 & 2 & 2 & 2 & 2 & 2 & 2 & 2 & 2 \\ 
  rep.month & 1 & 1 & 1 & 1 & 1 & 1 & 1 & 1 & 1 \\ 
  settlement.date & 2015/10/24 & 2015/10/24 & 2015/10/24 & 2015/10/24 & 2015/10/24 & 2015/10/24 & 2015/10/24 & 2015/10/24 & 2015/10/24 \\ 
  settlement.year & 4 & 4 & 4 & 4 & 4 & 4 & 4 & 4 & 4 \\ 
  dev.year & 1 & 2 & 3 & 4 & 5 & 6 & 7 & 8 & 9 \\ 
  calendar.year &  2 &  3 &  4 &  5 &  6 &  7 &  8 &  9 & 10 \\ 
  open & 1 & 1 & 1 & 0 & 0 & 0 & 0 & 0 & 0 \\ 
  settlement & 0 & 0 & 1 & 1 & 1 & 1 & 1 & 1 & 1 \\ 
  payment & 1 & 1 & 1 & 0 & 0 & 0 & 0 & 0 & 0 \\ 
  size & 1030.9341 & 1353.2233 &  517.9971 & 0 & 0 & 0 & 0 & 0 & 0 \\ 
\bottomrule
\end{tabular}}
\caption{Overview of one annualized claim simulating along the (baseline) working principles of the simulation machine. The payment sizes are aggregated per development year since reporting.\label{tab:overviewsimulationmachine}}
\end{table}
\end{document}